\documentclass[12pt]{iopart}

\usepackage{hyperref}
\usepackage{graphicx}
\usepackage{amssymb}
\usepackage[dvipsnames]{xcolor}
\usepackage{mathrsfs}

\newcommand{\nn}{\nonumber}
\newcommand{\cg}{\textnormal{\textsl{g}}}
\newcommand{\prt}[2]{\frac{\partial{#1}}{\partial{#2}}}
\usepackage[normalem]{ulem}

\begin{document}

\title[Non-Minimal Coupling of Magnetic Fields with Gravity in Schwarzschild Spacetime]{On the Non-Minimal Coupling of Magnetic Fields with Gravity in Schwarzschild Spacetime}

\author{Kumar Ravi$^{1}$\footnote{Corresponding author}, Petar Pavlovi\'c$^{2}$, Andrey Saveliev$^{3,4}$}
$^{1}$Ramakrishna Mission Vivekananda Educational and Research Institute, Belur Math, Howrah, West Bengal, India\\
$^{2}$Institute for Cosmology and Philosophy of Nature, Trg svetog Florijana 16, Križevci, Croatia \\
$^{3}$Immanuel Kant Baltic Federal University, Ul.~A.~Nevskogo 14, Kaliningrad, Russia \\
$^{4}$Lomonosov Moscow State University, GSP-1, Leninskiye Gory 1-52, Moscow, Russia
\ead{cimplyravi@gmail.com}

\begin{abstract}
In this work we study the effects of non-minimal coupling between electromagnetism and gravity, which are motivated by quantum effects such as vacuum polarization. We investigate the modification of both asymptotically dipole and uniform magnetic fields around a Schwarzschild black hole that come as the result of non-minimal coupling. While in both cases the magnetic field gets enhanced or suppressed with respect to the case of minimal coupling, depending on the sign of non-minimal coupling parameter, in the case of a background uniform magnetic field the direction of the magnetic field also alters in the vicinity of the black hole horizon. We have discussed the possible astrophysical and cosmological sources for which the vacuum polarization may be at play, while also discussing the observational effects, in particular the possibility of synchrotron radiation from the vicinity of a black hole. We conclude that such observations could be used to constrain the value of the non-minimal coupling parameter.
\end{abstract}

\section{Introduction}
The description of electromagnetism on macroscopic scales in terms of Maxwell equations represents one of the most successful, well established and also oldest field theory. The standard description of electromagnetic fields in the presence of gravitational fields, which is of interest in cosmology and astrophysics, therefore assumes a direct mathematical generalization of Maxwell's equations on curved spacetime \cite{Tsagas:2004kv}. On the other hand, the effect of electromagnetic fields on spacetime is described by the Einstein equation in the standard approach, where the stress-energy tensor of electromagnetic fields bends the spacetime in the same fashion as any other source of energy density. This type of dependence between gravity and electromagnetism is called minimal, since the field equations for this case are derived by a simple addition of gravitational and electromagnetic part of the Lagrangian, without any cross terms. There are, however, strong reasons to expect that this picture should be changed for very strong gravitational fields, where the non-minimal coupling -- the presence of Lagrangian terms containing direct contractions between gravitational and electromagnetic tensors -- should arise. First of all, we could expect that at sufficiently high energies gravity and electromagnetism will get united and described by a single field, similar to what already has been found for electricity and magnetism and electromagnetism and weak nuclear interaction. Since in this regime the gravitational and electromagnetic sector could change into each other and would manifest a deep interconnection, the investigation of non-minimal coupling between electromagnetism and gravity can serve as an effective model for such effects. 

There is also a more concrete reason for consideration of non-minimal coupling in regimes of strong gravitational fields. In \cite{Drummond:1979pp} the effect of quantum electrodynamics (QED) vacuum polarization onto curved spacetime was studied for the case of a photon propagating in vacuum and it was found that it leads to a non-minimal coupling between gravity and electromagnetism. This can be understood as coming from the transition of a photon into an electron/positron pair and the consequent tidal influences of the spacetime geometry along the characteristic Compton wavelength \cite{Drummond:1979pp, Pavlovic:2018idi}. In the one-loop approach it was demonstrated \cite{Drummond:1979pp} that the vacuum polarization effect thus leads to the effective Lagrangian density of the form
\begin{equation}
 \mathcal{L}=\frac{R}{\kappa} + \frac{1}{2}F^{\mu\nu}F_{\mu \nu} + \frac{1}{2}\mathcal{R}^{\mu\nu\rho\sigma} F_{\mu \nu}F_{\rho \sigma}+
 \mathcal{L}_{\rm matter}\,,
 \label{lagrangian}
\end{equation}
where $\kappa = 8 \pi G/c^{4}$ (from now on we will set $c=G=1$), $R$ is the Ricci scalar, $F^{\mu \nu}$ is the Maxwell tensor obeying $F^{\mu \nu}=\nabla^{\mu}A^{\nu} - \nabla^{\nu}A^{\mu}$, where $\nabla_{\mu}$ is the covariant derivative, and $\mathcal{L}_{\rm matter}$ is the Lagrangian of neutral matter. The effects of the vacuum polarization are contained in the tensor $\mathcal{R}^{\mu\nu\rho\sigma}$ (not to be confused with the Riemann tensor $R^{\mu \nu \rho \sigma}$) which is defined as %
\begin{eqnarray}
\mathcal{R}^{\mu\nu\rho\sigma} &\equiv \frac{q_{1}}{2}(g^{\mu\rho}g^{\nu \sigma} - g^{\mu\sigma}g^{\nu\rho})R \nn \\ 
&+ \frac{q_{2}}{2}(R^{\mu\rho} g^{\nu\sigma} - R^{\mu\sigma} g^{\nu\rho} + R^{\nu \sigma}g^{\mu\rho} - R^{\nu\rho}g^{\mu\sigma}) \nn \\
&+ q_{3}R^{\mu\nu\rho\sigma}\,, 
\label{oneloop}
\end{eqnarray}
where $q_{1}$, $q_{2}$ and $q_{3}$ are the coupling constants, and $R^{\mu \nu }$ is, as usual, the Ricci tensor and $R^{\mu\nu\rho\sigma}$ is the Riemann tensor. Physical consequences of such types of non-minimal coupling were investigated in various settings for a long time \cite{1971PhLA...37..331P, Prasanna:1973xv, Balakin:2005fu, Chu:2010tc, Savchenko:2018pdr, Cano:2021hje}.\\

In \cite{Pavlovic:2018idi} it was proposed that the signatures of non-minimal coupling between electromagnetism and gravity could, in principle, be observed or at least constrained by studying the magnetic fields around the event horizons of black holes and that the same effect could be used for constraining the sizes of primordial black holes. With this aim, the problem of modification of magnetic fields due to the vacuum polarization effect around a Schwarzschild black hole was for the first time studied in \cite{Pavlovic:2018idi} and the consequences for the orbits of charged particles around such black holes were for the first time studied in \cite{Pavlovic:2019rim}. In this work we try to further develop the investigation of this topic. We study the uniform magnetic field configuration, which has not been analysed yet in this context, and also improve and critically discuss the previously studied dipole magnetic field case. We then investigate the motion in the equatorial plane by using the effective potential and demonstrate that most of the conclusions that can be reached regarding the effects of the non-minimal coupling on the trajectories and scattering of the charged particles can be simply understood with the help of the effective potential study. 

This paper is organized as follows: In Sec.~\ref{sec:NMC} we briefly review the theory of non-minimal coupling, while in Sec.~\ref{sec:NMCSS} we present the analytical (minimal coupling) and numerical solutions (non-minimal coupling) of the Maxwell equations in Schwarzschild spacetime for the cases (i) of a static and asymptotically dipole magnetic field is being placed at the origin of the Schwarzschild black hole, and (ii) of a Schwarzschild black hole being placed in a static and asymptotically uniform magnetic field. Sec.~\ref{sec:MotionEqPlaneEffPot} is dedicated to the study of the motion of a charged test particle in the equatorial plane of the Schwarzschild spacetime in the effective potential formalism for both of these magnetic field configurations. In Sec.~\ref{sec:PartCollAcc} we briefly explore the energetics, collision and the possibility of acceleration of charged test particles near the event horizon when a Schwarzschild black hole is placed in a static and asymptotically uniform magnetic field. Subsequently, in Sec.~\ref{sec:Applications}, we present possible astrophysical and cosmological scenarios for which the considerations of vacuum polarization and hence the applications of non-minimal coupling might be important. Also, in this section we consider the possible observational signatures, before concluding this work with a discussion of its results and future scopes in Sec.~\ref{sec:Conclusions}.

\section{Non-Minimal Coupling} \label{sec:NMC}

There are, of course, numerous ways in which the non-minimal coupling between gravitational and electromagnetic sectors can be achieved, such as coupling between curvature tensors of different rank and order with vector potentials, Maxwell tensors and their contractions, etc.~\cite{1971PhLA...37..331P, Prasanna:1973xv, Balakin:2005fu, Bamba:2008ja, Dereli:2011hu, Dereli:2011mk, Sert:2015ykz, Sert:2018mls}. Many of those options are, however, completely arbitrary and lacking an additional motivation, and some of them lead to the violation of important physical principles, such as gauge invariance. We will focus on the type of non-minimal coupling discussed in Sec.~\ref{sec:NMC} and considered in \cite{Drummond:1979pp} since (i) it is motivated by the quantum-electrodynamical one-loop corrections in curved spacetime, (ii) it involves the contraction of all fundamental electromagnetic and curvature tensors to the leading order, and (iii) it preserves gauge invariance. When the variational procedure is performed on the Lagrangian (\ref{lagrangian}), we obtain the following equations of motion:
\begin{equation}
\nabla_{\mu}\left(F^{\mu\nu} + \mathcal{R}^{\mu\nu\rho\sigma}F_{\rho\sigma}\right) = 0\,.
\label{nonminEq}
\end{equation}

In this work we assume the coupling constants $q_{1},q_{2}, q_{3}$ to be phenomenological constants which should be constrained based on observations. In \cite{Drummond:1979pp} the authors calculated the values of those constants for the simplest case of a photon in vacuum on curved spacetime. Since it is not simple to see how this result can be generalized for macroscopic magnetic fields of arbitrary configurations, our approach appears to be completely justified. For a more detailed discussion of this issue see Sec.~\ref{sec:Applications} of this work. One should note that the freedom in the choice of these constants does not violate the gauge invariance, i.e.~the equations will stay gauge-invariant regardless of the value of the coupling parameters. To see this clearly one should observe that the requirement of gauge invariance was used in the computations presented in \cite{Drummond:1979pp} initially in order to determine that the part of the effective action describing the effects of the virtual electron loops depends on $F_{\mu \nu}$ and not on $A_{\mu}$ directly. The only three possible terms leading to non-minimal coupling which respect this condition, are gauge-invariant and which are linear in curvature invariants and $F_{\mu \nu}$ are the ones appearing in equation (\ref{oneloop}) multiplying the coefficients $q_{1},q_{2}, q_{3}$ \cite{Deser:1974cz}. As this is the general result, it will not be affected by any change of value in the coupling coefficients. In accord with this, further calculation of their specific values for the case of photon in vacuum presented in \cite{Drummond:1979pp} was not based on the requirement of gauge invariance, but was performed by comparing the action coming from the virtual electron loops with the weak-gravitational-field limit. 

The discussed non-minimal coupling will in general lead to the violation of equivalence principle of general relativity. The equivalence principle is essentially based on a possibility of making a local identification between an arbitrary spacetime manifold and the flat spacetime of special relativity, in a sufficiently small neighborhood around a point of the manifold. That is to say, the equivalence principle assumes that there is a tangent Minkowskian space around every point of the spacetime. However, when the effect of vacuum polarization is introduced, it becomes impossible to make the very transition to a flat spacetime around some point for the system of interest. Due to its quantum nature, manifested in its transition into a virtual electron/positron pair, a photon will acquire a characteristic length (corresponding to the Compton length) and due to the coupling with the gravitation field, it will be affected by tidal effects along this length. Since it, therefore, does not simply behave like a ``point-like" object, it is not possible to make a choice of a reference frame where the effects coming from its length would vanish, and where the propagation of a photon would be simply represented by a photon on a flat spacetime. As a general consequence, when non-minimal coupling coming from the vacuum polarization is introduced, it may be impossible to reduce the description of physics of such a spacetime to the principles of special relativity locally. A concrete consequence of such considerations was already demonstrated in \cite{Drummond:1979pp}. Namely, in specific reference frames photons in a gravitational field can travel with speeds greater than unity.

If one identifies $q_{1}=1/3,\,q_{2}=-1$ and 
$q_{3}=1$ then the tensor $\mathcal{R}^{\mu\nu\rho\sigma}$ turns out to be the Weyl tensor \cite{Balakin:2005xi}. There is an active field of research which studies the non-minimal coupling of photon to Weyl tensor \cite{Vagnozzi:2022moj,Jana:2021lqe,Cho:1997vg,Cai:1998ij,Prasanna:2003ix,Zhang:2021hit} and which found that photons are then subject to birefringence and apparent superluminal motion. These consequences demand a modification/correction of the background metric (actually two effective metrics for two polarizations of photons). In the present work we are only focused on the situation where the background metric is still valid and we explore the motion/energetics of charged particles. The exploration of situations where non-minimal coupling of both the magnetic field (and therefore the motion/energetics of charged particles therein) and photons could be investigated together is an interesting project for the future.

\section{Non-Minimal Coupling in a Schwarzschild Spacetime} \label{sec:NMCSS}

At this point we need to stress the most important physical approximations which we use in our study, which were also assumed in earlier works \cite{Pavlovic:2018idi, Pavlovic:2019rim}. 

We first assume that magnetic fields are weak enough for their effect on the spacetime being negligible, such that the latter can therefore still be described by the Schwarzschild metric. While not necessarily asymptotically flat, there are some exact metrics where, unlike here, the magnetic fields are not treated as just test fields (see \cite{Karas:2014paa} and references therein). Finding the complete solutions to both the non-minimally coupled Maxwell and Einstein equations is, however, of significant interest, and should be considered in future research. Such full solutions of the non-minimal problem, describing both electromagnetic and gravitational sector, could be of physical importance in the very early Universe, when the non-minimal coupling effects and the resulting magnetic fields might have very strong (see discussion in Sec.~\ref{sec:Conclusions}). The approximation of a weak field and a negligible feedback on the spacetime we use here should, however, be justified for all known astrophysical systems (see detailed discussion in \cite{Pavlovic:2018idi,Aliev:1989wx}).

The next crucial assumption we made is to ignore the electric fields and consider only the magnetic part of the electromagnetic tensor. This is justified because of the very high value of the conductivity of the Universe, causing the electric fields to be completely insignificant -- this assumption therefore represents one of the standard elements in studies of astrophysical electromagnetism. On the other hand, magnetic fields probably exist on all scales of the observable Universe (see \cite{ PhysRevD.37.2743,Beck:2000dc,Giovannini:2003yn,Balakin:2005xi,Bernet:2008qp,Subramanian:2009fu,Han:2009ts,Widrow:2011hs,DuNe,Subramanian:2015lua,Vachaspati:2020blt,Batista:2021rgm} and references therein). This fact makes magnetic fields an excellent candidate for investigation of the effect of non-minimal coupling between electromagnetism and gravity. We therefore do not need to consider a special mechanism for the creation of magnetic fields around black holes, since every black hole will naturally be immersed at least within the galactic magnetic field. 

The physical picture we consider here is therefore the following: we assume the existence of a magnetic field, going to its asymptotical dependence far away from the black hole -- where the spacetime is approximately flat -- while being influenced by the black hole spacetime near the black hole horizon. The magnetic field we treat here is thus not the magnetic field of the black hole itself, but an external magnetic field which can be galactic or belong to some other source, which we immerse in the Schwarzschild spacetime. 

The Schwarzschild metric in a coordinate system adapted to spherical symmetry $(t,\,r,\,\theta,\,\phi)$ is
\begin{equation}
{\rm d}s^{2} = -\left(1 - \frac{2M}{r}\right){\rm d}t^{2} + \left(1 - \frac{2M}{r}\right)^{-1}{\rm d}r^{2} + r^{2}\left({\rm d}\theta^{2} + \sin^{2}{\theta}{\rm d}\phi^{2} \right)\,,
\end{equation}
where the mass $M$ is the source of gravity. The Schwarzschild solution of the Einstein field equation is a unique solution and is obtained using the assumptions of spherical symmetry, asymptotic flatness, and a vacuum outside the spherical object of mass $M$. On top of that, this solution turns out to be static. In accordance with the already discussed weak field approximation we will assume that the effects of magnetic fields on the spacetime are negligible, and, as a direct consequence of this assumption, that the deviations from the Schwarzschild vacuum solution are negligible when the magnetic fields are introduced on that spacetime.

We assume a magnetic field, expressed in the local Lorentz frame (LLF) \cite{GinzburgOzernoi1965,prasanna1977charged}, 
\begin{eqnarray}
F_{\theta\phi}^{\rm LLF} &=& B_{r} =\frac{2\mu\cos\theta}{r^{3}}\xi(r)\,,\nn\\
F_{\phi{r}}^{\rm LLF} &=& B_{\theta} =
\frac{\mu\sin\theta}{r^{3}}\psi(r)\,,\nn\\
F_{r\theta}^{\rm LLF} &=& B_{\phi} = 0\,,
\label{Maxten0}
\end{eqnarray}
where $\mu$ is the magnetic moment and the functions $\xi(r)$ and $\psi(r)$ take into account the effects of curved spacetime on the magnetic field. We will later see that $\xi(r)$ and $\psi(r)$ may be obtained from a system of second order linear differential equations, and so we get two linearly independent solutions: one corresponding to a dipole magnetic field, while the other corresponds to a uniform magnetic field. Both of these cases are the most basic models for studying magnetic fields. For instance, the galactic magnetic field is modeled by the dipole magnetic field, so the discussed model should be suitable for description of black holes immersed in the galactic magnetic field.

After doing the transformation from local Lorentz frame, the components of Maxwell tensor in Schwarzschild spacetime are
\begin{eqnarray}
F_{\theta\phi} &=& \frac{2\mu\sin\theta\cos\theta}{r} \xi(r)\,,\nn\\
F_{\phi{r}} &=& \frac{\mu\sin^{2}\theta}{r^{2}\sqrt{1 - \frac{2M}{r}}}\psi(r)\,,\nn\\
F_{r\theta} &=& 0\,.
\label{Maxten1}
\end{eqnarray}
As for any vacuum solution of the Einstein field equation, the Ricci scalar ($R$) and Ricci tensor ($R_{\mu\nu}$) are zero, such that Eq.~(\ref{nonminEq}) reduces to
\begin{equation}
\nabla_{\mu}\left(F^{\mu\nu} + 
q_{3}R^{\mu\nu\rho\sigma}F_{\rho\sigma}\right) = 0\,.
\label{MaxEq1}
\end{equation}
The Maxwell tensor components are also subject to another Maxwell equation, namely
\begin{equation}
\nabla_{[\lambda}F_{\mu\nu]} = 0\,.
\label{MaxEq2}
\end{equation}
After substituting the Maxwell tensor components from Eq.~(\ref{Maxten1}) into Eq.~(\ref{MaxEq1}) and into Eq.~(\ref{MaxEq2}), we get
\begin{equation}
\frac{{\rm d}}{{\rm d}r}\left[\left(1-\frac{2 M q_{3}}{r^{3}} \right)\frac{\left(1 - \frac{2 M}{r}\right)^{1/2}}{r^{2}}\,\psi(r)\right] + \left(1 + \frac{4 M q_{3}}{r^{3}} \right)\frac{2\xi{\left(r \right)}}{r^{3}} = 0\,,
\label{Maxeq1a}
\end{equation}
and
\begin{equation}
\frac{{\rm d}}{{\rm d} r}\left[\frac{\xi(r)}{r}\right] + \frac{\psi{\left(r \right)}}{r^{2} \sqrt{1- \frac{2 M}{r}}}=0\,,
\label{Maxeq2a}
\end{equation}
respectively. These equations can easily be decoupled, resulting in
\begin{eqnarray}
&\frac{{\rm d}}{{\rm d}r}\left[\left(1 - \frac{2Mq_{3}}{r^{3}} \right)\left(1 - \frac{2M}{r} \right)\frac{{\rm d}}{{\rm d}r}\left(\frac{\xi(r)}{r}\right) \right] \nn\\
&- \left(1 + \frac{4Mq_{3}}{r^{3}} \right)\frac{2\xi(r)}{r^{3}}=0\,.
\label{Dcoup1}
\end{eqnarray}
Here, we do not write the decoupled equation for $\psi(r)$, not just due to its complicated form, but also due to the fact that it will not be used in our computations. In fact, it is much easier to use the solutions of Eq.~(\ref{Dcoup1}) and insert them into Eq.~(\ref{Maxeq2a}), thus obtaining the solutions for $\psi(r)$, rather than solving another differential equation from the beginning. 

Before getting into the solutions of Eq.~(\ref{Maxeq1a}) and Eq.~(\ref{Maxeq2a}), i.e.~for the case of non-minimal coupling, we briefly review the available solutions for the case of minimal coupling.

\subsection{Minimal Coupling: A Brief Review of the Analytical Solutions}

As expected, with $q_{3} = 0$ the Eqs.~(\ref{Maxeq1a}) and (\ref{Maxeq2a}) reduce to the corresponding equations of \cite{GinzburgOzernoi1965} and \cite{prasanna1977charged}. While for the case of minimal coupling (i.e.~$q_{3}=0$), getting analytical solutions is straightforward with the application of the Frobenius series solution technique for a second order ordinary linear homogeneous equation, we could not carry out the same for $q_{3} \neq 0$. The solutions for the case of minimal coupling are \cite{GinzburgOzernoi1965,prasanna1977charged,1983ApJ...265.1036W}
\begin{eqnarray}
&&\xi^{\rm d}(r) = -\frac{3r^{3}}{8M^{3}}\left[\ln\left(1- \frac{2M}{r}\right)
+\frac{2M}{r}\left(1 + \frac{M}{r}\right)\right],\nn\\
&&\psi^{\rm d}(r) = \frac{3r^{2}}{4M^{2}}\Bigg[1 + \left(1- 
\frac{2M}{r}\right)^{-1} + \frac{r}{M}\ln\left(1- 
\frac{2M}{r}\right)\Bigg]\sqrt{1- \frac{2M}{r}}\,,
\label{dpSol0}
\end{eqnarray}
and
\begin{eqnarray}
&&\xi^{\rm u}(r) = \frac{B_{0}}{2\mu}r^{3}\,,\nn\\
&&\psi^{\rm u}(r) = -\frac{B_{0}}{\mu}r^{3}\sqrt{1-\frac{2M}{r}}\,,
\label{unSol0}
\end{eqnarray}
where $B_{0}$ is the constant of integration and can be interpreted as the magnitude of the asymptotically uniform background magnetic field. We have used the superscripts `d' and `u' for the dipole solution and for the uniform solution, respectively. It can easily be checked that for the dipole solution (Eq.~(\ref{dpSol0})), asymptotically for $r \rightarrow \infty$, the two functions behave as $\xi(r)\rightarrow1$ and $\psi(r)\rightarrow1$, such that Eq.~(\ref{Maxten0}) reduces to the familiar flat-spacetime solution. Therefore, in LLF the expressions for the asymptotically dipole magnetic field solution are
\begin{eqnarray}
B_{r}^{\rm d} &=& -\frac{3\mu\cos\theta}{4M^{3}}\left[\ln\left(1- \frac{2M}{r}\right)
+\frac{2M}{r}\left(1 + \frac{M}{r}\right)\right],\nn\\
B_{\theta}^{\rm d} &=& \frac{3\mu\sin\theta}{4M^{2}r}\Bigg[1 + \left(1-
\frac{2M}{r}\right)^{-1} + \frac{r}{M}\ln\left(1- 
\frac{2M}{r}\right)\Bigg]\,,\nn\\
\label{MagnF1}
\end{eqnarray}
while for the asymptotically uniform magnetic field we obtain
\begin{eqnarray}
B_{r}^{\rm u} &=& B_{0}\cos\theta\,,\nn\\
B_{\theta}^{\rm u} &=& -B_{0}\sin\theta\sqrt{1 - \frac{2M}{r}}\,.
\label{MagnF2}
\end{eqnarray}
It can be seen that solutions (\ref{MagnF1}) contain a singularity at $r=2M$, which is a coordinate singularity associated with crossing the horizon in the given coordinates. This naturally raises the question whether this singularity implies the violation of the assumption of weak magnetic fields and suggests that magnetic fields significantly affect the metrics, thus making the whole treatment problematic. Indeed, the very black hole horizon and the points in the very vicinity of it cannot be covered by the analysis provided here, since they depart from the regime of weak magnetic fields. However, for distances that are still near the horizon (for instance around the value of $r=1.5 \times r_{0}$) magnetic fields will still be weak and will not violate the initial assumption of not affecting the metrics. This happens because the singularity contained in (\ref{MagnF1}) is actually a ``mild" logarithmic singularity, leading to significant increase of the field only in the very small neighborhood around $r_{0}$. This can, for instance, clearly be seen in Figs.~\ref{Brdp} and \ref{Bthdp} which demonstrate that the solutions remain small for the astrophysically interesting region around the horizon. As we are studying the propagation of charged particles around the black hole, this behavior at the very horizon is not of concern, but one should note that our treatment cannot be applied there.

We should note that $\xi(r)$ and $\psi(r)$ basically represent the ratios of magnetic field components in a curved (here Schwarzschild) spacetime to those in the flat spacetime (e.g.~the quantities on vertical axis in the Fig.~1 of \cite{Pavlovic:2018idi}) and these ratios are precisely what we are interested in. 
Now we will proceed with the investigations for the non-minimal coupling scenario.

\subsection{Non-Minimal Coupling: Numerical Solutions}

In order to numerically solve the coupled system of differential equations (\ref{Maxeq1a}) and (\ref{Maxeq2a}) we need two initial/boundary conditions, namely $\xi(r_{0})$ and $\psi(r_{0})$ at some $r_{0}$. For a chosen value of $q_{3}$, we can choose large enough $r_{0}$ such that $4Mq_{3}/r_{0}^{3} \ll 1$ and then safely assume the solutions there, at $r_{0}$, to be equal to those of minimal coupling scenario with excellent precision. Therefore, in order to obtain the asymptotically dipole solution numerically we can choose the initial conditions to be compatible with Eq.~(\ref{dpSol0}), and for the asymptotically uniform solution with Eq.~(\ref{unSol0}). For instance, $r_{0}\simeq 100 \times 2M$, $\xi(r_{0})\simeq1$ and $\psi(r_{0})\simeq1$ are valid initial conditions for seeking the dipole solution, whereas for the uniform solution $\xi(r_{0})\simeq {\rm const.}\times r_{0}^{3}/2$ and $\psi(r_{0})\simeq -{\rm const.}\times r_{0}^{3}\sqrt{1-2M/r_{0}}$ are a reasonable approach. We define a dimensionless form of the coupling parameter as $\tilde{q} = q_{3}/(2M)^{2}$.

\begin{figure}[t]
\centering
\includegraphics[scale=0.45]{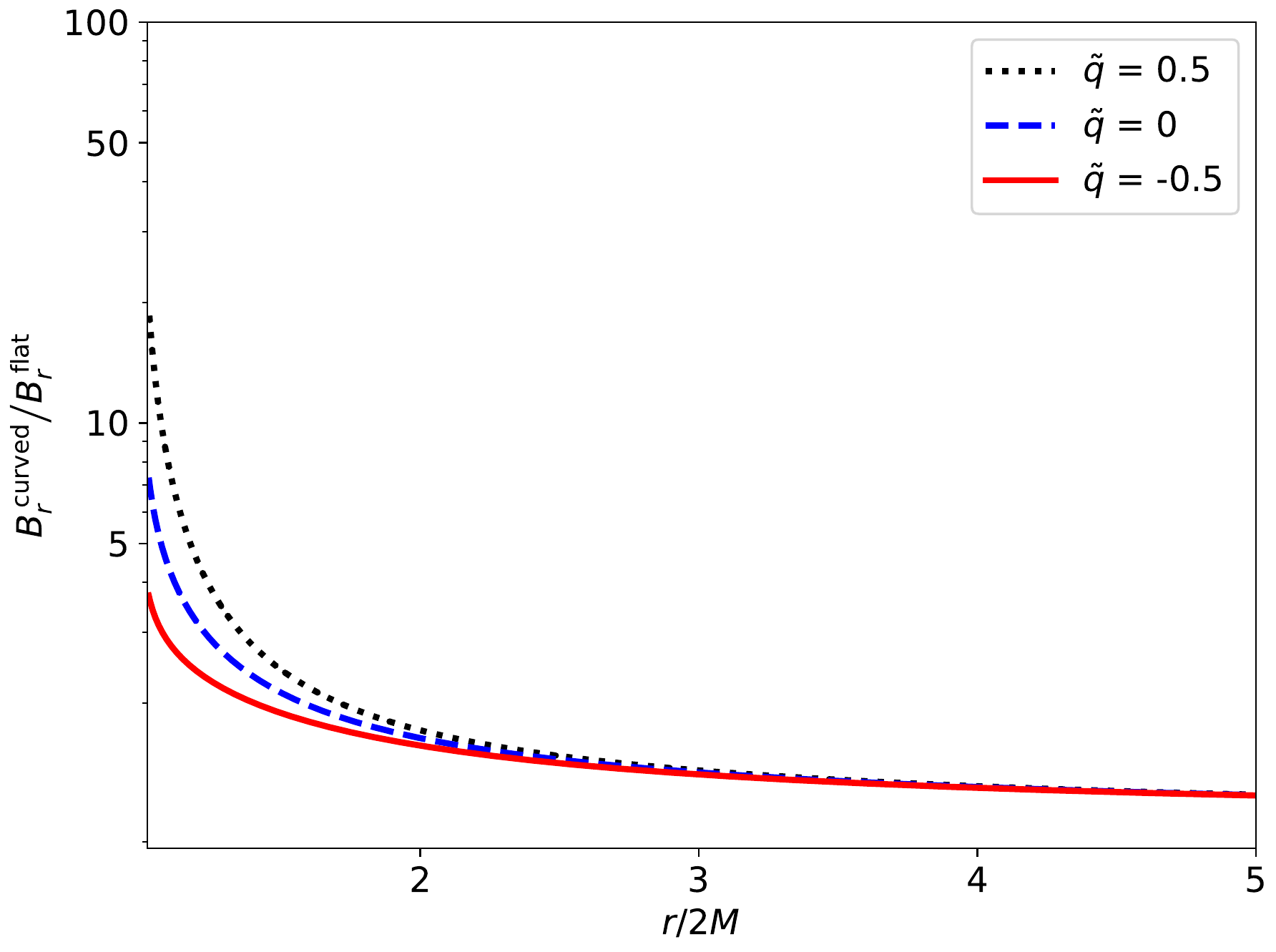}
\caption{The radial component of a dipole magnetic field in Schwarzschild spacetime ($B_{r}^{\rm curved}$) scaled by the same quantity in flat spacetime ($B_{r}^{\rm flat}$) as a function of $r$ for different values of $\tilde{q}$. The enhancement/suppression of the magnetic field component near the horizon for positive/negative values of $\tilde{q}$ compared to the case of minimal coupling ($\tilde{q}=0$) can be seen.}
\label{Brdp}
\end{figure}

\begin{figure}[t]
\centering
\includegraphics[scale=0.45]{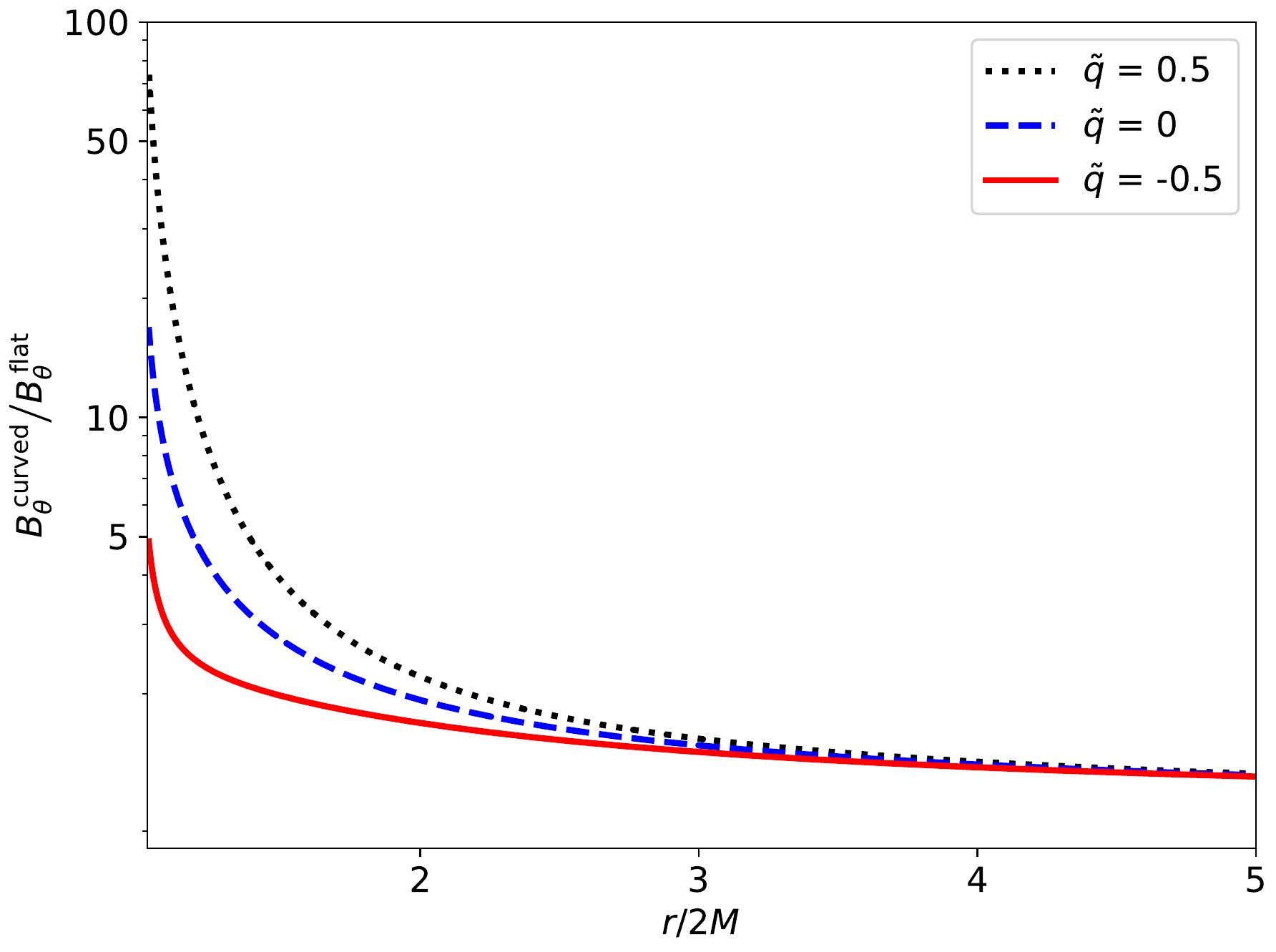}
\caption{The azimuthal component of a dipole magnetic field in Schwarzschild spacetime ($B_{\theta}^{\rm curved}$) scaled by the same quantity in flat spacetime ($B_{\theta}^{\rm flat}$) as a function of $r$ for different values of $\tilde{q}$. The enhancement/suppression of the magnetic field component near the horizon for positive/negative values of $\tilde{q}$ compared to the case of minimal coupling ($\tilde{q}=0$) can be seen.}
\label{Bthdp}
\end{figure}

Following the numerical recipe described above, we have obtained the magnetic field solutions for the non-minimal coupling case. In Figs.~\ref{Brdp} and \ref{Bthdp} we have plotted the ratio of the radial and azimuthal component of the asymptotically dipole magnetic field, modified by gravity to that of a dipole magnetic field in flat spacetime, both for the case of minimal and non-minimal coupling scenarios, i.e.~$\xi(r)$ and $\psi(r)$, respectively. We can see from these figures that for a positive (negative) coupling constant $\tilde{q}$ there is an enhancement (suppression) of magnetic field present near the horizon. 

We should compare the results of \cite{Pavlovic:2018idi,Pavlovic:2019rim} with the present work. Although the main conclusions and the qualitative behavior is the same in these previous works and the one in this study, there are still some quantitative differences present. Though far from the event horizon, in the asymptotically flat region, both this presentation and that of \cite{Pavlovic:2018idi,Pavlovic:2019rim} do not differ significantly, and both have the same typical dipole form ($B_{r}(r,\theta) = 2{\rm const.}\times\cos\theta/r^{3}$ and $B_{\theta}(r,\theta) = {\rm const.}\times\sin\theta/r^{3}$), however they differ near the event horizon. The radial dependence of both the components $B_{r}(r,\theta)$ and $B_{\theta}(r,\theta)$ in \cite{Pavlovic:2018idi,Pavlovic:2019rim}, namely $B_{\rm rad}(r)$, is identical, but here we have different radial dependences for these two components, namely $\xi(r)$ and $\psi(r)$. These differences come from the different mathematical treatment. In \cite{Pavlovic:2018idi,Pavlovic:2019rim} the geometry of the magnetic field configuration was prescribed \textit{a priori}, assuming that $B_{\theta}(r,\theta)=\tan \theta B_{r}(r,\theta)/2$. This type of assumption regarding the $B_{\theta}$ and $B_{r}$ is based on the configuration satisfying the second Maxwell equation on flat spacetime, viz.~$\nabla \times \mathbf{B}=0$. The modifications of the magnetic field were then inspected using Eq.~(\ref{MaxEq1}), while no further reference was made with respect to the other Maxwell equation, Eq.~(\ref{MaxEq2}), since it is not modified by the presence of the non-minimal coupling. In this work, however, no $\textit{a priori}$ assumption regarding the field configuration, and thus the relationship between $B_{\theta}$ and $B_{r}$, has been made, and both Maxwell equations are solved simultaneously, making the treatment complete and self-consistent, thus improving the mathematical analysis. The predicted enhancement/suppression for a given $\tilde{q}$ in \cite{Pavlovic:2018idi,Pavlovic:2019rim} is larger than what the present work predicts about the component $B_{r}$, whereas for $B_{\theta}$ component it is, in fact, smaller.

\begin{figure}[t]
\centering
\includegraphics[scale=0.45]{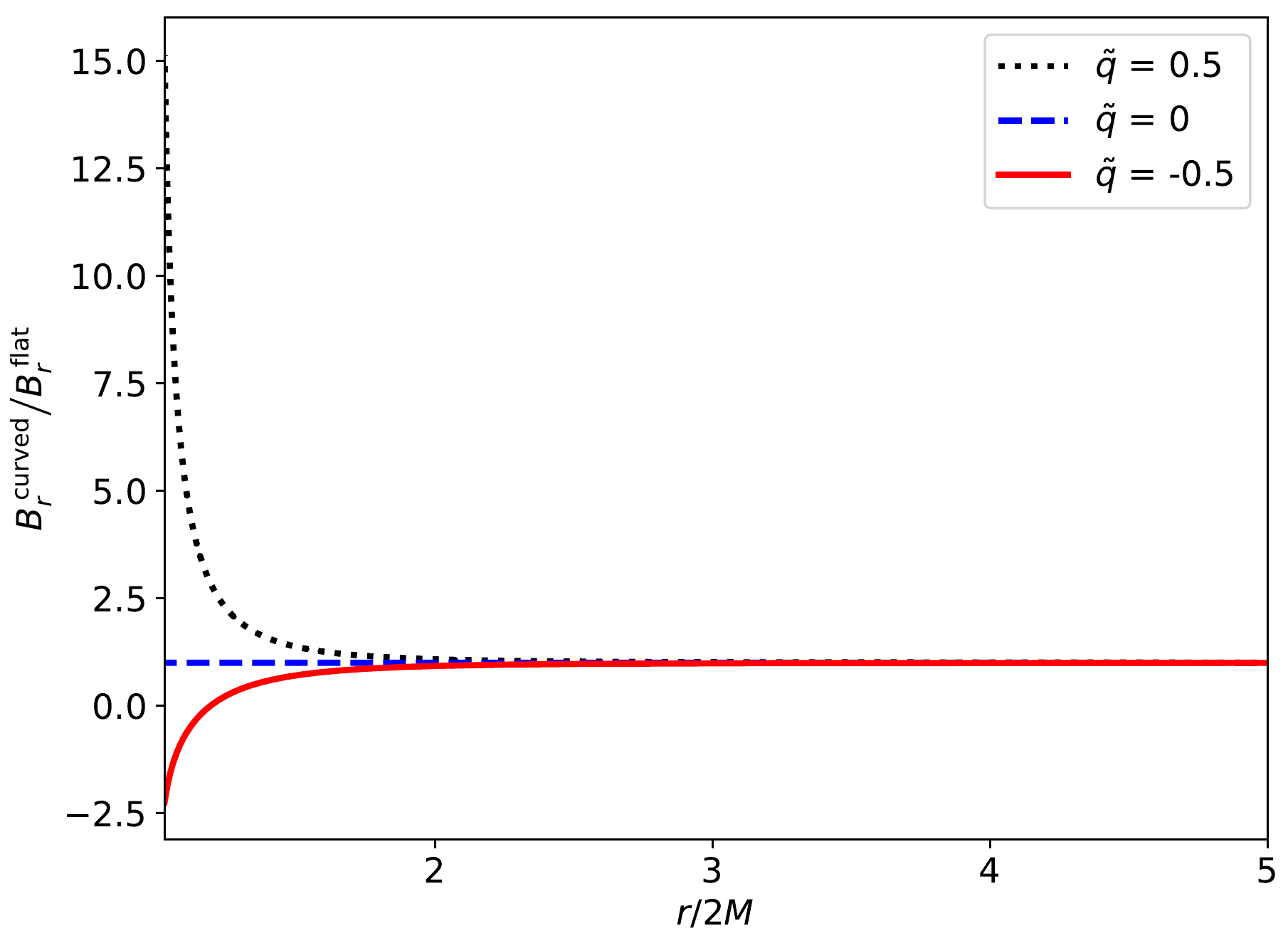}
\caption{ The radial component of the uniform magnetic field in Schwarzschild spacetime scaled by the same quantity in flat spacetime as a function of $r$ for different values of $\tilde{q}$. For $\tilde{q} = 0.5$ the enhancement of the magnetic field component near the horizon with respect to the case of minimal coupling ($\tilde{q}=0$) can be seen, while for $\tilde{q} = -0.5$ we observe the change in sign and magnitude of the field component near the horizon.}
\label{Brun}
\end{figure}

\begin{figure}[t]
\centering
\includegraphics[scale=0.45]{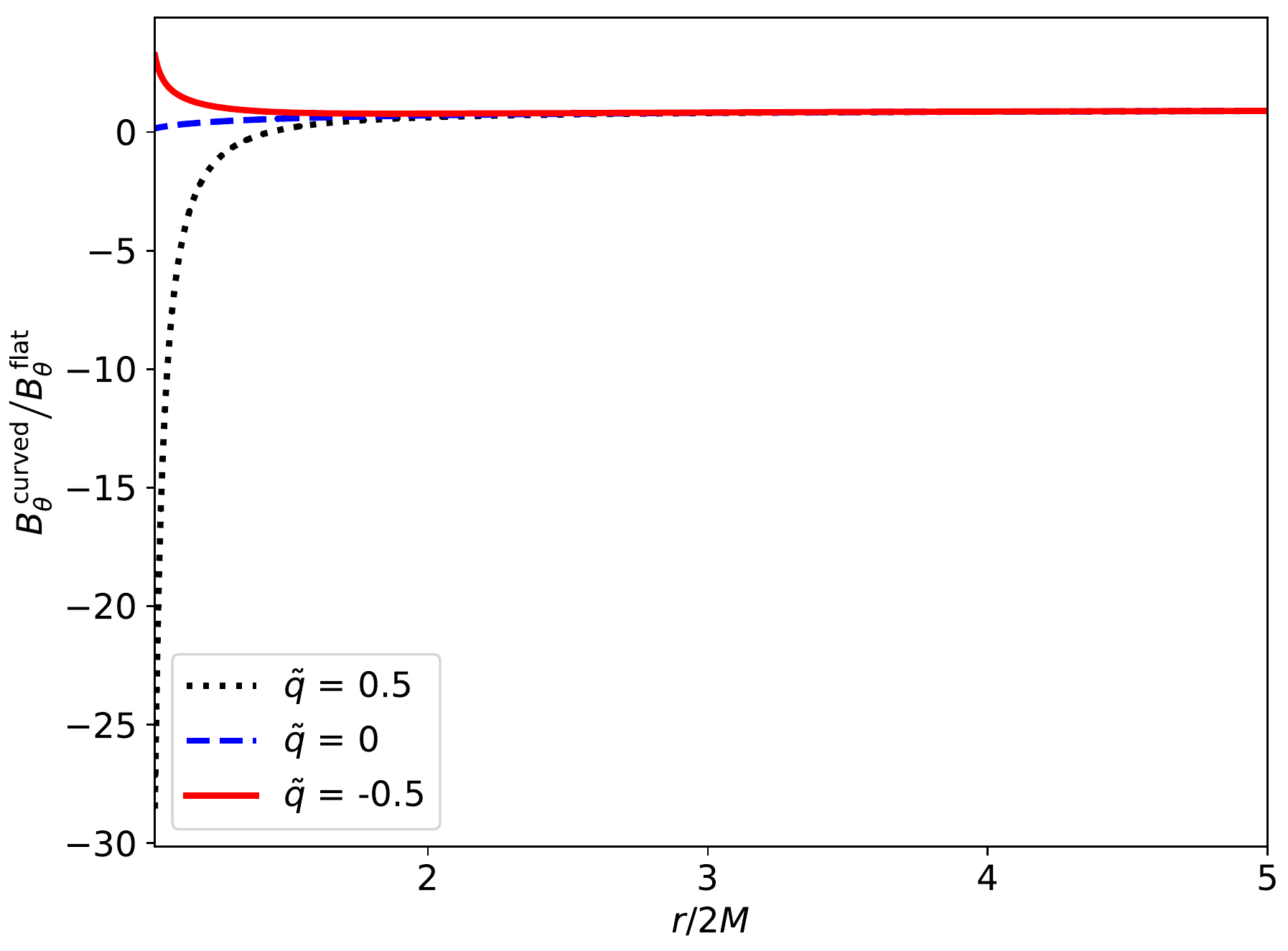}
\caption{The azimuthal component of the uniform magnetic field in Schwarzschild spacetime scaled by the same quantity in flat spacetime as a function of $r$ for different values of $\tilde{q}$. For $\tilde{q}=-0.5$ the enhancement of the magnetic field component near the horizon with respect to the case of minimal coupling ($\tilde{q}=0$) can be seen, while for $\tilde{q}=0.5$ we observe the change in sign and magnitude of the field component near the horizon.}
\label{Bthun}
\end{figure}

\begin{figure}[t]
\centering
\includegraphics[scale=0.6]{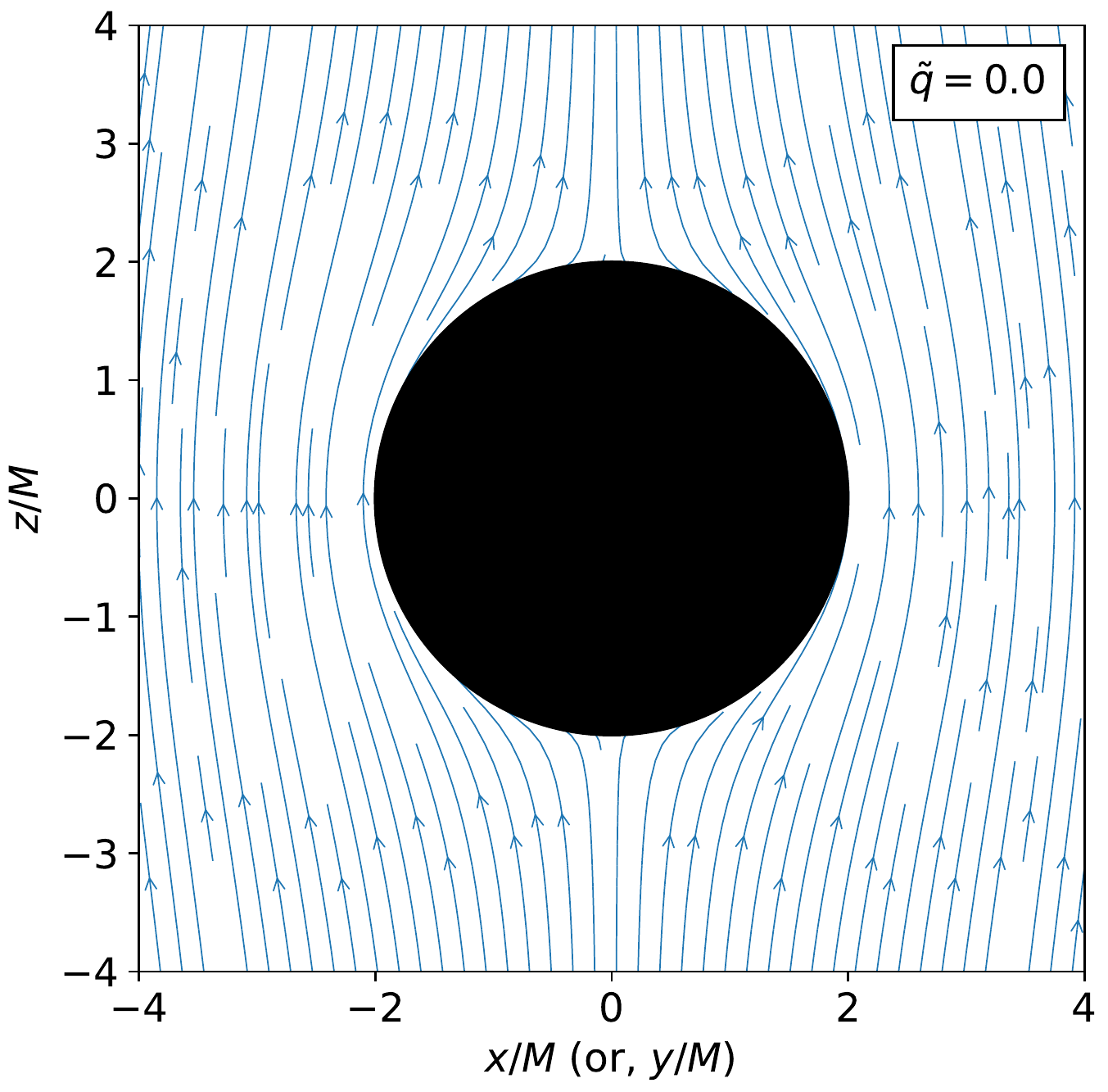}
\caption{An asymptotically uniform magnetic field near the Schwarzschild horizon for the minimal coupling scenario.}
\label{Fieldlines0}
\end{figure}

\begin{figure}[t]
\centering
\includegraphics[scale=0.6]{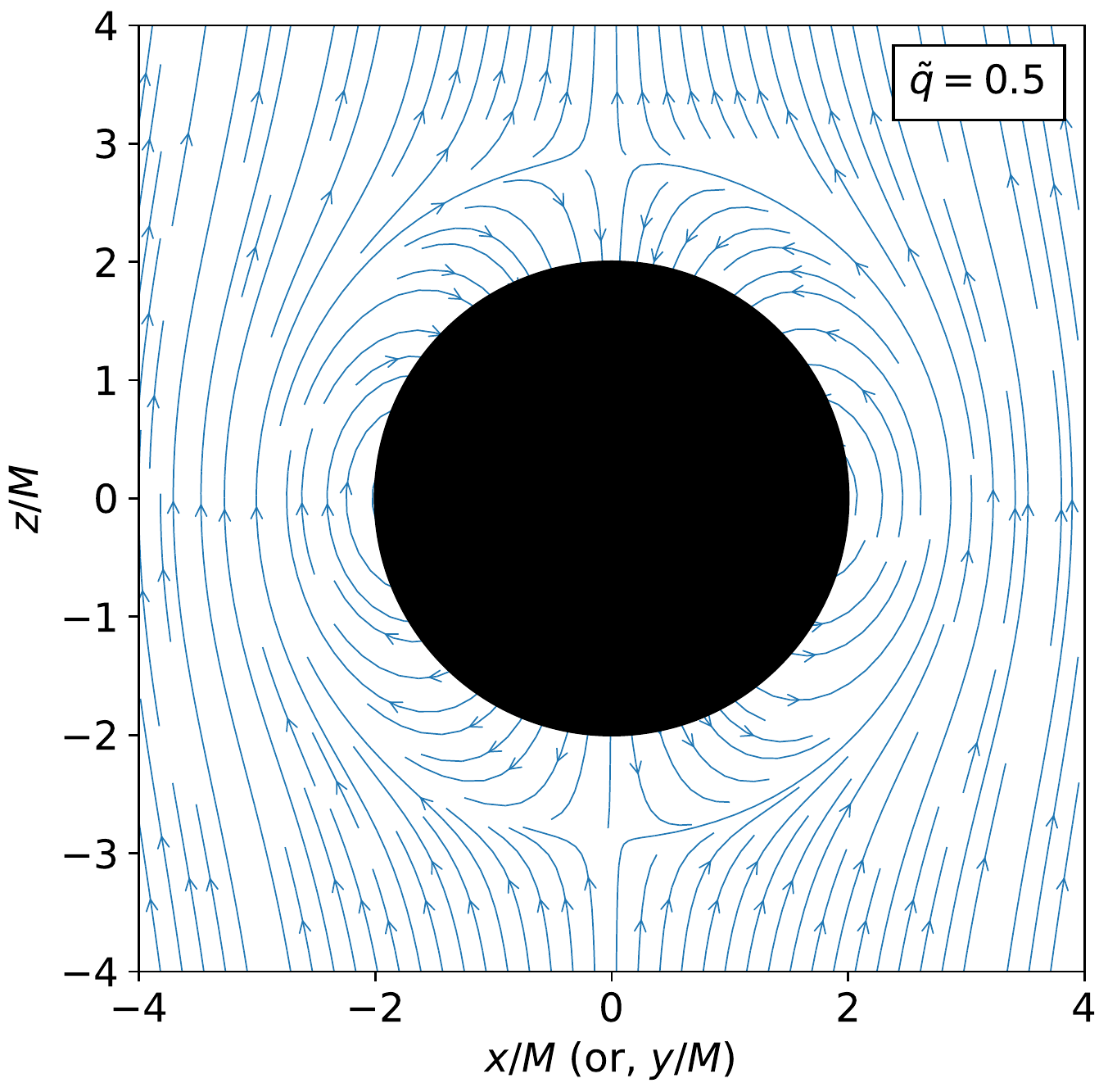}
\caption{An asymptotically uniform magnetic field near the Schwarzschild horizon for the non-minimal coupling parameter $\tilde{q}=0.5$.}
\label{Fieldlines_p}
\end{figure}

\begin{figure}[t]
\centering
\includegraphics[scale=0.6]{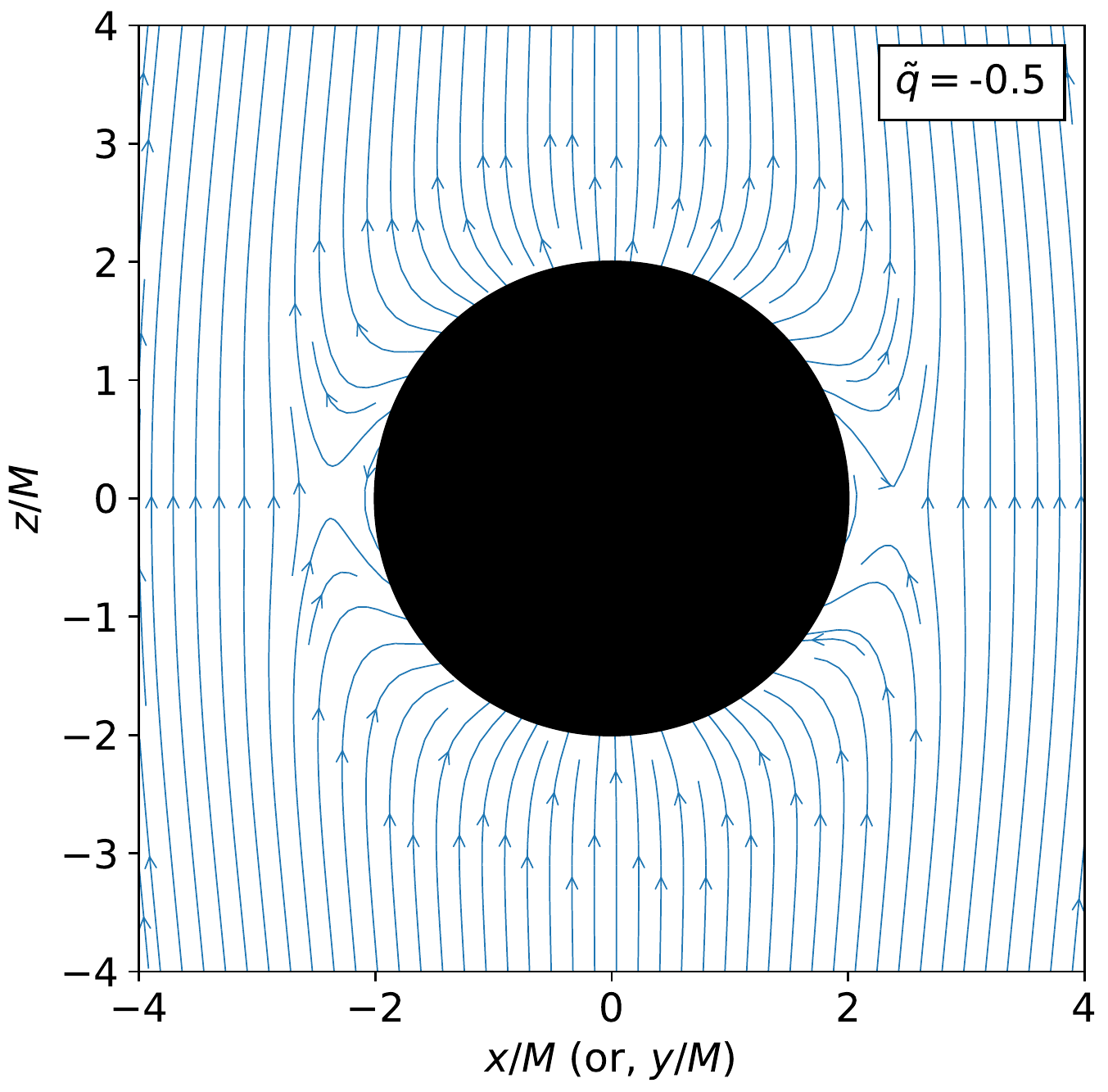}
\caption{An asymptotically uniform magnetic field near the Schwarzschild horizon for the non-minimal coupling parameter $\tilde{q}=-0.5$.}
\label{Fieldlines_n}
\end{figure}

The modification of an asymptotically uniform magnetic field near a Schwarzschild black hole in a minimal coupling scenario has been widely studied in the literature (see \cite{Frolov:2010mi} and references therein). As expected, the components of asymptotically uniform magnetic field get enhanced/suppressed near the event horizon depending on the sign of the coupling constant $\tilde{q}$ (see Figs.~\ref{Brun} and \ref{Bthun}). There are two points to note here: (i) in minimal coupling scenario the radial component, $B_{r}$, remains unaffected, whereas in the case of non-minimal coupling both the components, $B_{r}$ and $B_{\theta}$, are affected; (ii) for $\tilde{q}\neq0$ the direction of magnetic field changes near the horizon -- for positive values of $\tilde{q}$ the direction switches to the opposite direction near the equatorial plane, whereas for negative values of $\tilde{q}$ the direction changes near the two poles (see Figs.~\ref{Fieldlines_p} and \ref{Fieldlines_n}; to contrast them from the case of minimal coupling scenario, see Fig.~\ref{Fieldlines0}). Therefore, in contrast to the case of a dipole field, an asymptotically uniform magnetic field is also modified qualitatively, apart from the quantitative enhancement/suppression both cases have in common. The manifestations of this qualitative change on the motion of a charged particle would be worth further exploration. But since in the following we study the motion in the equatorial plain, the effect of the change of direction near the poles (for $\tilde{q}>0$) is irrelevant.

As can be seen in Fig.~\ref{Fieldlines_n}, the asymptotically uniform magnetic field is being repelled by the black hole in the vicinity of the Schwarzschild horizon, i.e.~similar to the Meissner effect in superconductors. This Meissner-like effect near the black hole horizon vis-a-vis its undermining effects for the efficiency of the Blandford-Znajek mechanism for jet formations in active galactic nuclei (AGNs), gamma-ray bursts (GRBs), galactic black hole binaries, etc., has been widely studied (see, for example, \cite{Bicak:2006hs, Komissarov:2007rc, Penna:2014aza} and references therein). In this work we are only considering the vacuum solution, such that our results are not opposed to the Blandford-Znajek mechanism which requires a plasma-filled magnetosphere around the black hole, where this Meissner-like effect have been found to be absent by many researchers.

\section{Motion in the Equatorial Plane: Effective Potential} \label{sec:MotionEqPlaneEffPot}

Due to the symmetry of the problem we are investigating, the 4-potential can be assumed to have the form \cite{Petterson:1975sg}
\begin{equation}
A_{\mu} = \left(0,0,0,A_{\phi}(r,\theta)\right)\,,
\end{equation}
while the Lagrangian for the motion of a charged test particle of charge $Q$ and mass $m$ can be written as \cite{Preti_2004}
\begin{equation}
\mathscr{L} = \frac{1}{2} m \cg_{\mu\nu}\dot{x}^{\mu}\dot{x}^{\nu} + Q A_{\phi}\dot{\phi}\,,
\end{equation}
where $\cg_{\mu\nu}$ denotes the metric components and $\dot{x}^{\mu}$ is the 4-velocity. From this Lagrangian, apart from the equation of motion
\begin{equation}
\frac{{\rm d}^{2}x^{\mu}}{{\rm d}\tau^{2}} + \Gamma^{\mu}{}_{\rho\sigma}\frac{{\rm d}x^{\rho}}{{\rm d}\tau}\frac{{\rm d}x^{\sigma}}{{\rm d}\tau} = \frac{Q}{m}F^{\mu}{}_{\nu\,\rm}\,\frac{{\rm d}x^{\nu}}{{\rm d}\tau}\,,
\label{eom}
\end{equation}
the two conserved quantities
\begin{equation}
\prt{\mathscr{L}}{\dot{t}} = - m \left(1 -\frac{2M}{r}\right) \frac{{\rm d} t}{{\rm d}\tau}\equiv -E \label{eng}
\end{equation}
and
\begin{equation}
\prt{\mathscr{L}}{\dot{\phi}} = mr^{2}\sin^{2}\theta\frac{{\rm d}\phi}{{\rm d}\tau} + QA_{\phi}\equiv l
\label{amom}
\end{equation}
may also be obtained. Here we stress that $F^{\mu\nu}$ represents the total electromagnetic tensor which already includes the corrections to magnetic fields coming from the non-minimal coupling effect (and not only the ``background'' part of the field) because it is the solution of the non-minimally coupled Maxwell equation, as can already be seen from the Eqs.~(\ref{Maxten1}) and (\ref{MaxEq1}). The potential $A_{\mu}$ discussed here is defined with respect to this total field in the usual manner, i.e.~$F_{\mu \nu}= \partial_{\mu} A_{\nu} - \partial_{\nu} A_{\mu}$. $E$ and $l$ can be interpreted as the (effective) energy and angular momentum, respectively. Therefore, although in the considered equations the terms responsible for the non-minimal coupling are not explicitly present, they are \textit{de facto} implicitly contained in the field terms such as $A_{\mu}$. We define $\tilde{E}\equiv E/m$, $\tilde{l}\equiv l/m$ and $q \equiv Q/m$. From the equation of motion (\ref{eom}) and the conservation equations (\ref{eng}) and (\ref{amom}), we get
\begin{equation}
\tilde{E}^{2} = \left(\frac{{\rm d}r}{{\rm d}\tau}\right)^{2} + V_{\rm eff}^{2}
\end{equation}
for the motion in the equatorial plane, where
\begin{equation}
V_{\rm eff}^{2} = \left(1 - \frac{2M}{r}\right) \left[1 + \frac{1}{r^{2}}\left(\tilde{l} - qA_{\phi}(r)\right)^{2}\right]\,,
\label{Veff}
\end{equation}
and, furthermore, $A_{\phi}(r)$ is subject to the ordinary differential equation
\begin{equation}
 \frac{{\rm d} A_{\phi}}{{\rm d}r}= -\mu\left(1 -\frac{2M}{r}\right)^{-1/2} \frac{\psi}{r^{2}}\,,
 \label{aphi}
 \end{equation}
 where the non-minimal coupling enters through the modifications of the function $\psi$.
 
\subsection{Dipole Magnetic Field}

We should mention the available analytical expression for $A_{\phi}(r,\theta)$ for the case of minimal coupling scenario \cite{prasanna1977charged},
\begin{equation}
A_{\phi}= -\frac{3\mu\sin^{2}\theta}{8M^{3}}r^{2}\left[\ln\left(1-\frac{2M}{r}\right) + \frac{2M}{r}\left(1+\frac{M}{r}\right)\right]\,.
\label{aa}
\end{equation}
For the non-minimal coupling scenario $V_{\rm eff}^{2}$ can be obtained numerically by solving the system of Eqs.~(\ref{Maxeq1a}), (\ref{Maxeq2a}) and (\ref{aphi}). The proper initial/boundary conditions to be used is $A_{\phi}(r_{0})\simeq0$ when $r_{0} \rightarrow \infty$ (which is compatible with Eq.~(\ref{aa})) and, again, those for $\xi(r)$ and $\psi(r)$ are as discussed earlier. We introduce two dimensionless parameters, $\lambda \equiv \mu{q}/M^{2}$ and $H \equiv \tilde{l}/M$, as an aid to study the effective potential.

\subsubsection{Case 1: The Angular Momentum is Parallel to Magnetic Moment}

\begin{figure}[t]
\centering
\includegraphics[scale=0.45]{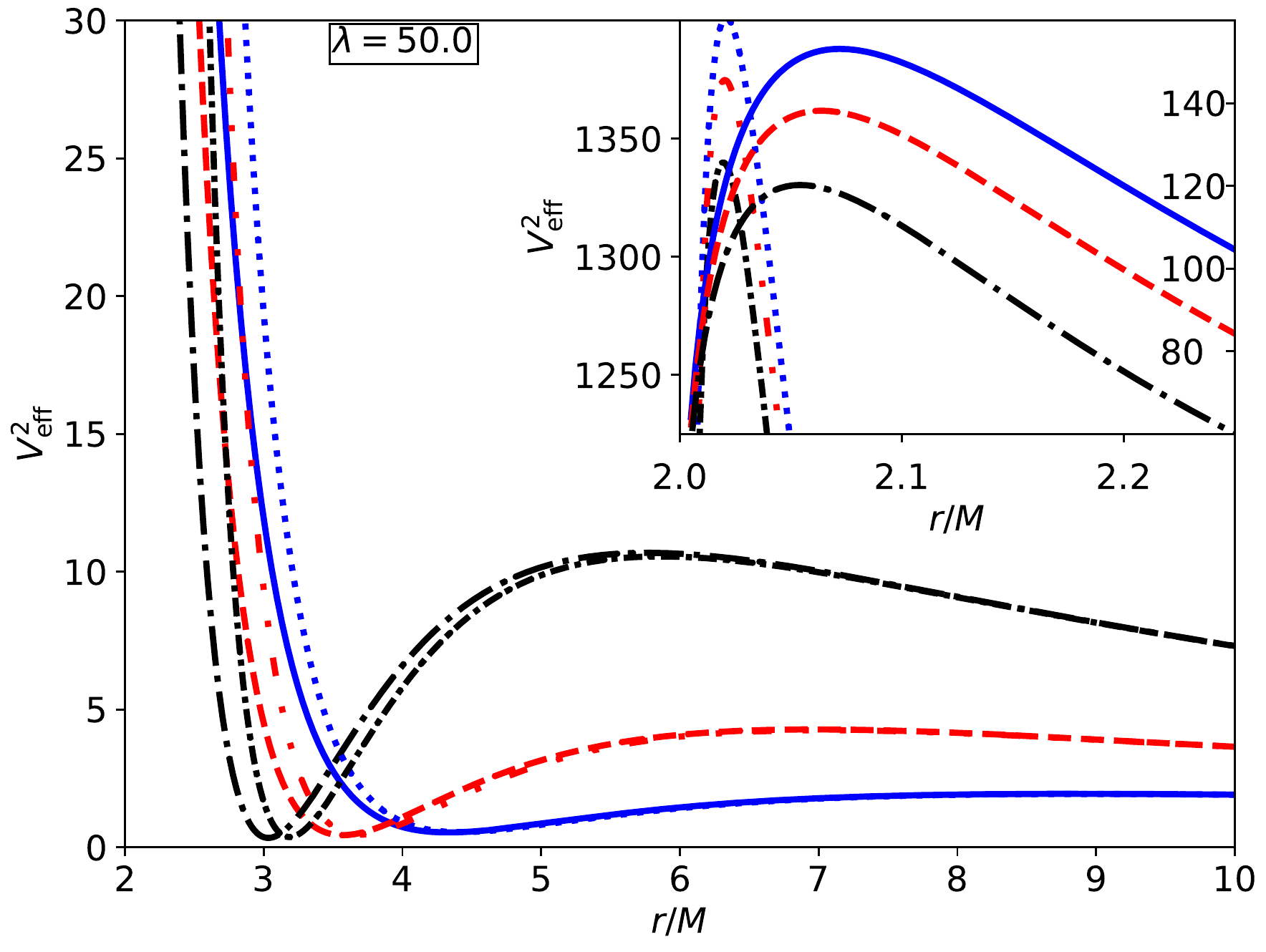}
\caption{The effective potential for the motion of a charged particle in the equatorial plane of the Schwarzschild spacetime in an asymptotically dipolar magnetic field, for $\lambda=50$. The colors blue, red and black correspond to the angular momentum parameter $H$ values 17.6954, 24.7736 and 34.4136, respectively. The {\it solid, dashed, dashdotted} curves are for $\tilde{q}=0.0$ and {\it dotted, dashdotdotted, densely dashdotted} curves are for $\tilde{q}=0.5$, respectively. Due to a lack of space we have presented these line-styles in the {\it figure legend} of Fig.~\ref{EPDqpHn} (the only differences are the signs of $H$ and $\tilde{q}$, while the magnitudes are the same). The values for $\lambda$ and $H$ are taken from \cite{prasanna1977charged}. In order to overcome the scaling difficulties, the peaks of the effective potentials are plotted in the inset where the left labeling of vertical axis is for $\tilde{q}=0.5$ and right labeling for $\tilde{q}=0$.}
\label{EPDqpHp}
\end{figure}

Let us first review some basic features of this setting in the minimal coupling case. In general, the feature-wise richest type of the effective potential has two maxima and two minima \cite{prasanna1977charged}. Between these two maxima a potential well (minimum) is present. The maxima nearest to the horizon are shown in the respective figure insets, while the minima farthest from the horizon are not shown, as they are not of significant physical interest. For a given value of $H$, increasing $\lambda$ amounts to a flattening-out of the potential well, raising the level of the first maximum, lowering the level of the outer maximum and shifting the first minimum outwards -- in case it exists. For a given $\lambda$, increasing $H$ amounts to narrowing the potential well, raising the level of the outer maxima, lowering the level of the inner maxima and shifting the first minimum inwards (see Figs.~\ref{EPDqpHp} and \ref{EPDqpHn}). So, if we focus on the case for which no outer maximum (and thus no potential well) initially exists for a given $\lambda$ and $H$, then with increasing $H$ at one point the outer maximum appears. Furthermore, its level increases with $H$, and thus the potential well becomes narrower and deeper. Finally, it reaches a maximum depth when both the maxima attain equal height. If $H$ is then further increased, the depth of the potential well starts decreasing as the inner maximum level decreases. Hence, increasing the value of $H$ further, the potential well will at some point again vanish as the inner maximum is vanishing (e.g.~$\lambda=50$, $H \sim 20-150$). 

In case of non-minimal coupling, a positive value of the coupling constant $\tilde{q}$ shifts the potential well outwards, does not affect the depth of the potential well significantly (just slightly reducing it) and raises the level of the first maximum significantly (see the inset of Fig.~\ref{EPDqpHp}). A negative value of $\tilde{q}$ shifts the potential well inwards and lowers the first minimum significantly (so it can affect the potential well, see Fig.~\ref{EPDqpHn}). Depending on the particle energy $\tilde{E}$, it will be subjected to scattering, a stable/unstable circular orbit, a bounded orbit with Larmor motion, or it may plunge into the black hole.

{\it \textbf{Scattering:}} A particle coming from infinity and having an energy smaller than either of the maxima will be scattered away back to infinity. When the inner maximum is the absolute maximum, depending on the sign of $\tilde{q}$, the spectrum (i.e.~the energy dependence) of the scattered particles will differ. For a positive $\tilde{q}$ particles with higher energies will be scattered away, while for a negative $\tilde{q}$ the scattered particles will have smaller energies than in the case of minimal coupling (see Figs.~\ref{EPDqpHp} and \ref{EPDqpHn}). When the outer maximum is the absolute maximum (e.g.~$\lambda = 30$, $H = 70.7816$ \cite{prasanna1977charged}), a nonzero $\tilde{q}$ will not affect the scattering significantly, as the outer maximum is not significantly influenced by the effect of vacuum polarization. 

{\it\textbf{Plunging into the black hole:}} A particle coming from infinity with an energy bigger than both of the maxima will plunge into the black hole. The effects of a nonzero $\tilde{q}$ on the spectrum of particles plunging into the black hole will be the opposite to what has been discussed regarding scattering.

{\it\textbf{Circular motion:}} A particle coming from infinity with energy equal to either of the maxima will have a unstable circular orbit. Therefore, when $\tilde{q}$ is positive (negative), it is possible for particles of respectively higher (lower) energies (in comparison to the case of minimal coupling) to have an unstable circular orbit. On the other hand, a particle having the energy which is equal to the first minimum and of local origin (i.e.~not coming from infinity) will have a stable circular orbit. This kind of circular orbit goes by the popular term ``innermost stable circular orbit'' ({\it isco}). In the absence of a magnetic field a charged particle cannot have a stable/unstable circular orbit below a radius of $6M$, while a magnetic field allows circular orbits well below this radius. Furthermore, in the non-minimal coupling case, with nonzero $\tilde{q}$, this radius gets additionally affected (it increases/decreases for positive/negative values of $\tilde{q}$, see Figs.~\ref{EPDqpHp} and \ref{EPDqpHn}). This will have implications for the modeling of accretion disks, as will be discussed later.

\par
\begin{figure}[t]
\centering
\includegraphics[scale=0.45]{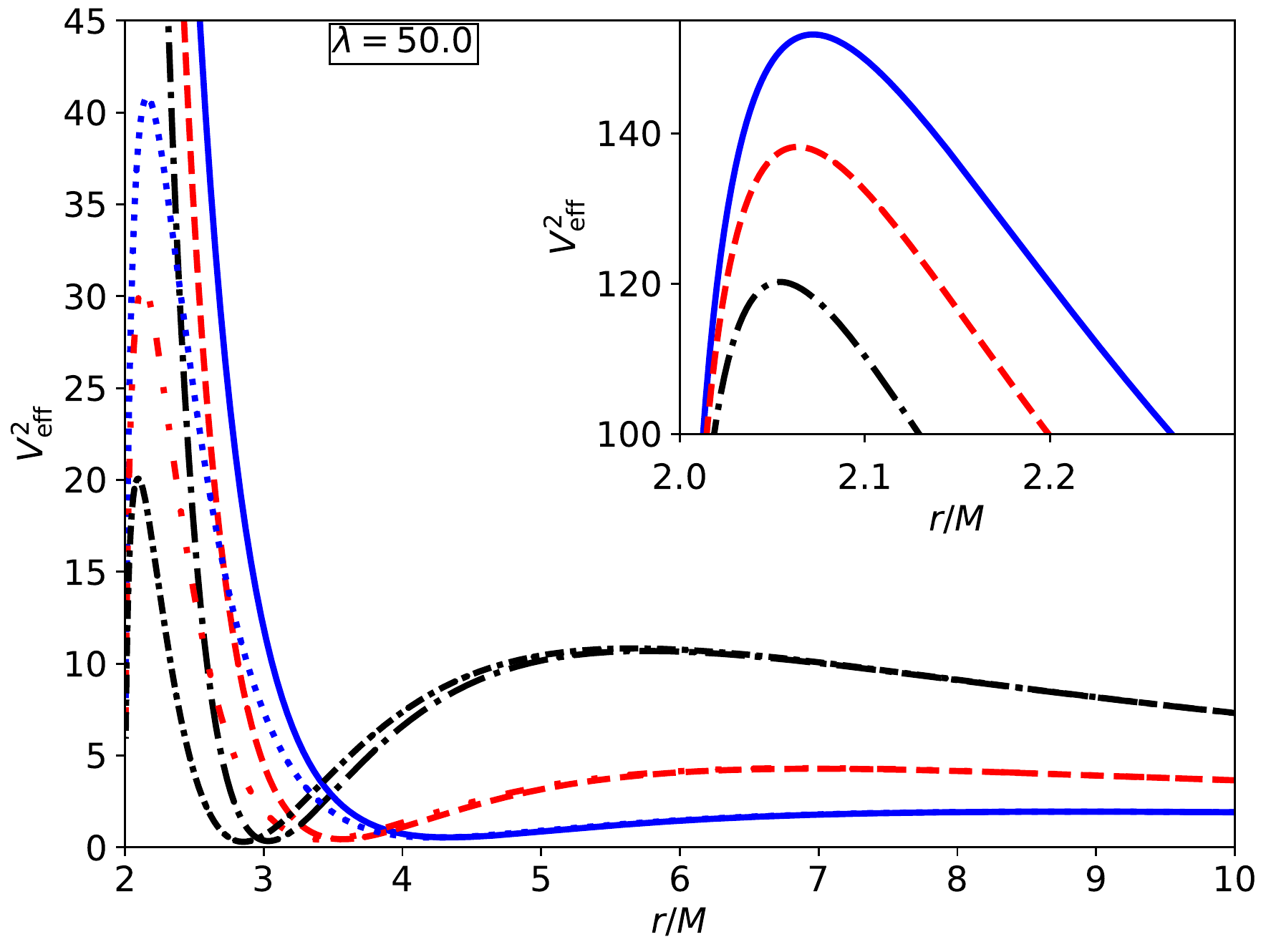}
\caption{Same as Fig.~\ref{EPDqpHp}, but for $\tilde{q}=-0.5$. The peaks of effective potentials for $\tilde{q} = 0.0$ are plotted in the inset. The decrease in the heights of the peaks for $\tilde{q} = -0.5$ in relation to the case of the minimal coupling can be seen.}
\label{EPDqnHp}
\end{figure}

\begin{figure}[t]
\centering
\includegraphics[scale=0.45]{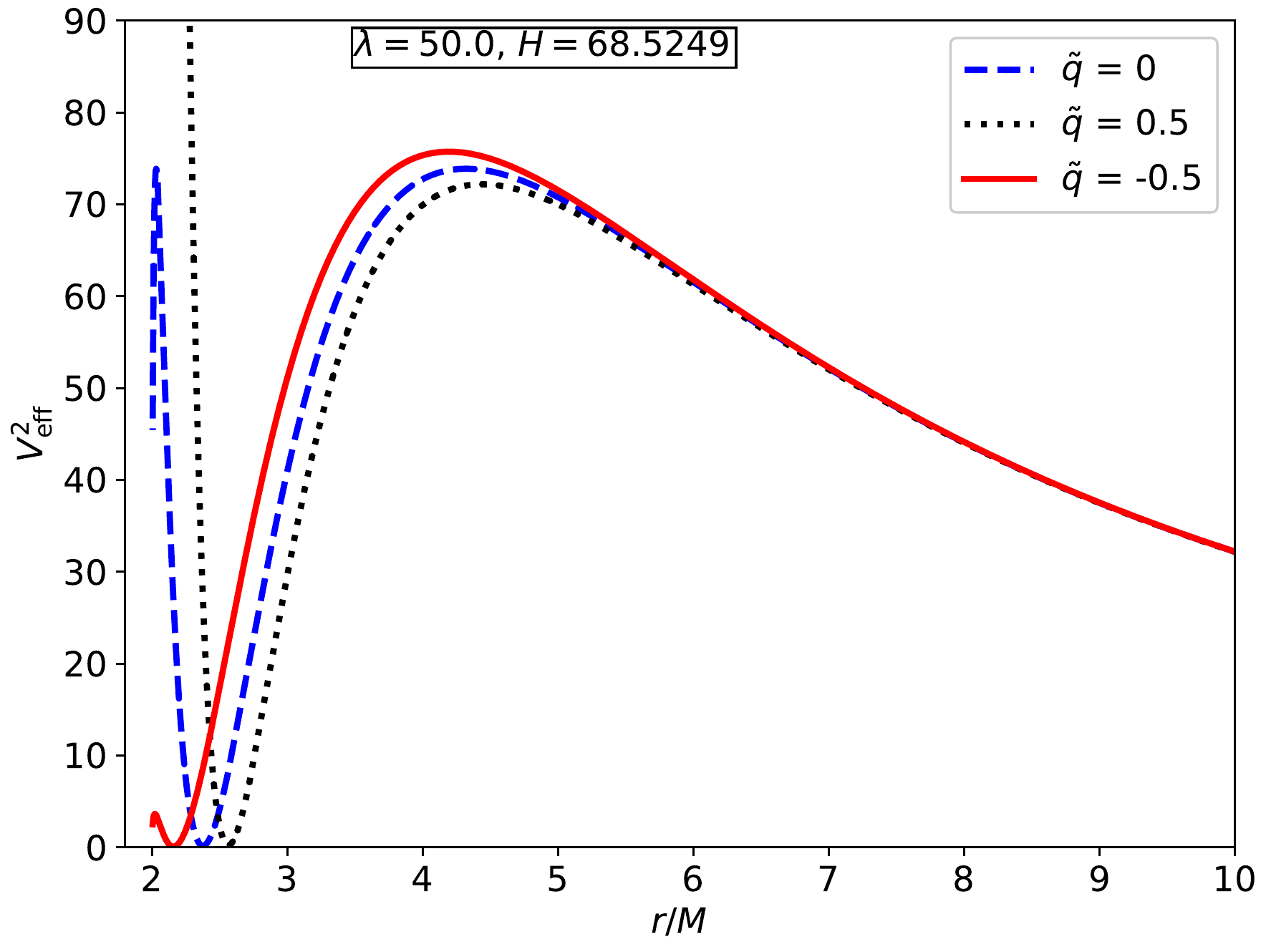}
\caption{The effective potential for the purpose of illustrating maximum depth potential (and hence maximal particle trapping). The used parameter values are $\lambda=50.0,\,H = 68.5249$ and the values of $\tilde{q}$ are shown in the legends. As can be seen, a negative coupling constant ($\tilde{q}$) significantly reduces the capacity of trapping charged particles near the horizon, while a positive $\tilde{q}$ does not affect it so much.}
\label{EPmaxtrap}
\end{figure}

 {\it\textbf{Particle trapping:}} A particle of local origin which has an energy smaller than both of the maxima will see two turning points and will be trapped in a gyrating (Larmor motion, see, for example, Figs.~5 to 9 of \cite{prasanna1977charged} and/or Figs.~1 to 3 of \cite{Pavlovic:2018idi}) bound orbit. Depending on the values of $\lambda$ and $H$ the two maxima can have the same height, such that the depth of the corresponding potential well will be maximum (e.g.~$\lambda= 50$,\,$H= 68.5249$). A positive value of $\tilde{q}$, apart from shifting the potential well outwards, hardly makes any difference in comparison with the minimal coupling scenario (see Fig.~\ref{EPmaxtrap}), while a negative $\tilde{q}$ significantly affects the potential well by modifying the first maximum and thus the trapping of particles (see Fig.~\ref{EPDqnHp}). A more negative value of $\tilde{q}$ can bring down the first maximum even lower with respect to the outer maximum and can hence make the potential well shallower (or can even wash it out completely (see Fig.~\ref{EPmaxtrap}). On the other hand, in the case when a negative $\tilde{q}$ does not make the potential well shallower (for certain combinations of values of $\lambda$, $H$, see e.g.~Fig.~\ref{EPDqnHp}), then -- as the potential well is in this case positioned closer to the event horizon -- this well is capable of holding more energetic particles trapped in itself in comparison to the minimal coupling scenario.

In all the settings discussed above the study of the spectrum of the scattered particles stands out as the most distinguished case for manifesting the difference between minimal and non-minimal coupling scenarios. With respect to the spectrum of the scattered particles, all other aspects of motion represent a less significant kind of probe that could serve as a signature for the existence of the effect of non-minimal coupling.

\subsubsection{Case 2: The Angular Momentum is Anti-Parallel to the Magnetic Moment}

\begin{figure}[t]
\centering
\includegraphics[scale=0.45]{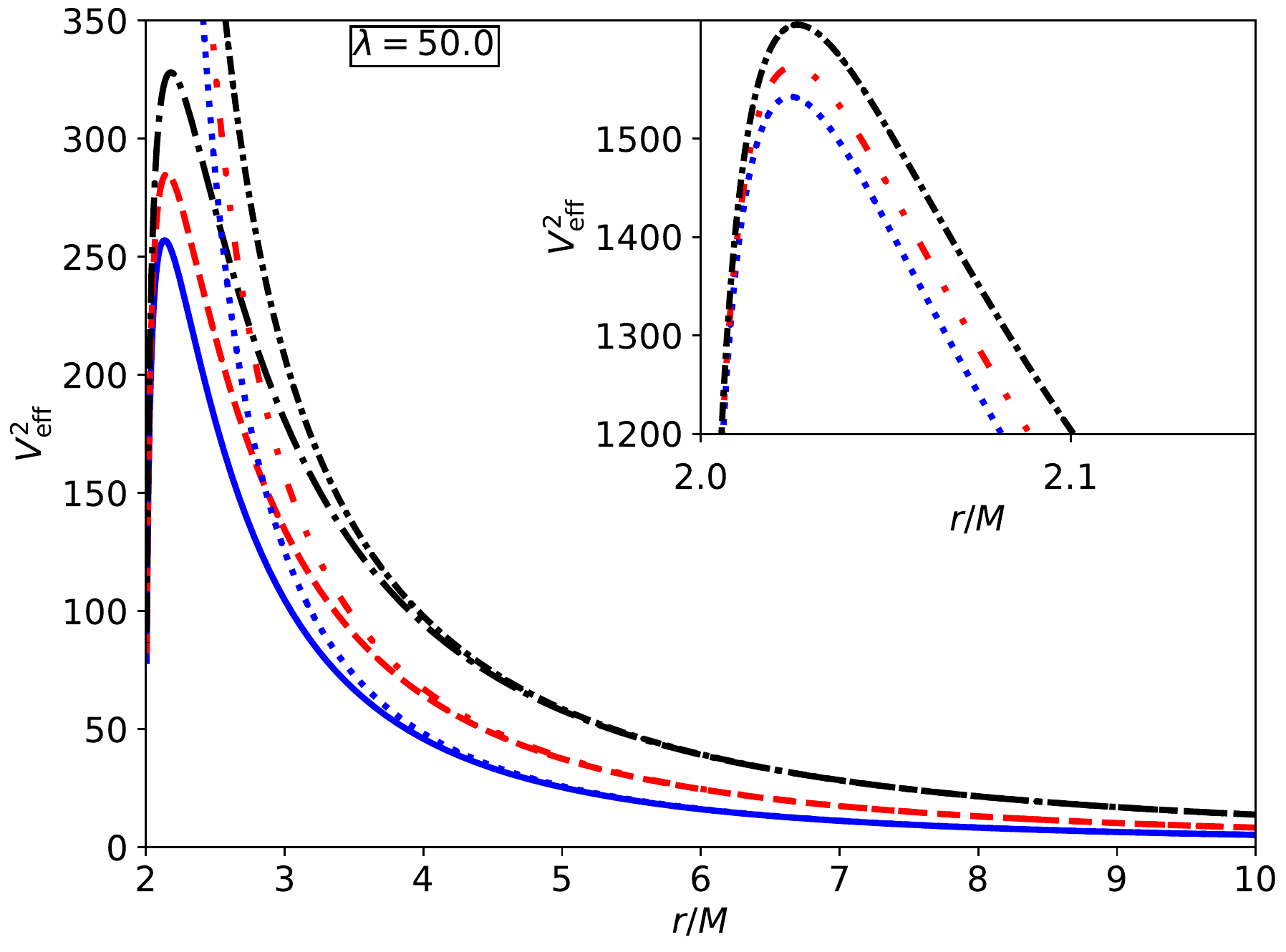}
\caption{Same as Fig.~\ref{EPDqpHp}, but here for the negative values of the angular momentum parameter ($H$) used there. The peaks for $\tilde{q} = 0.5$ are shown in the inset. An increase in the heights of the peaks for $\tilde{q} = 0.5$ with respect to the case of minimal coupling may be seen here.}
\label{EPDqpHn}
\end{figure}

\begin{figure}[t]
\centering
\includegraphics[scale=0.45]{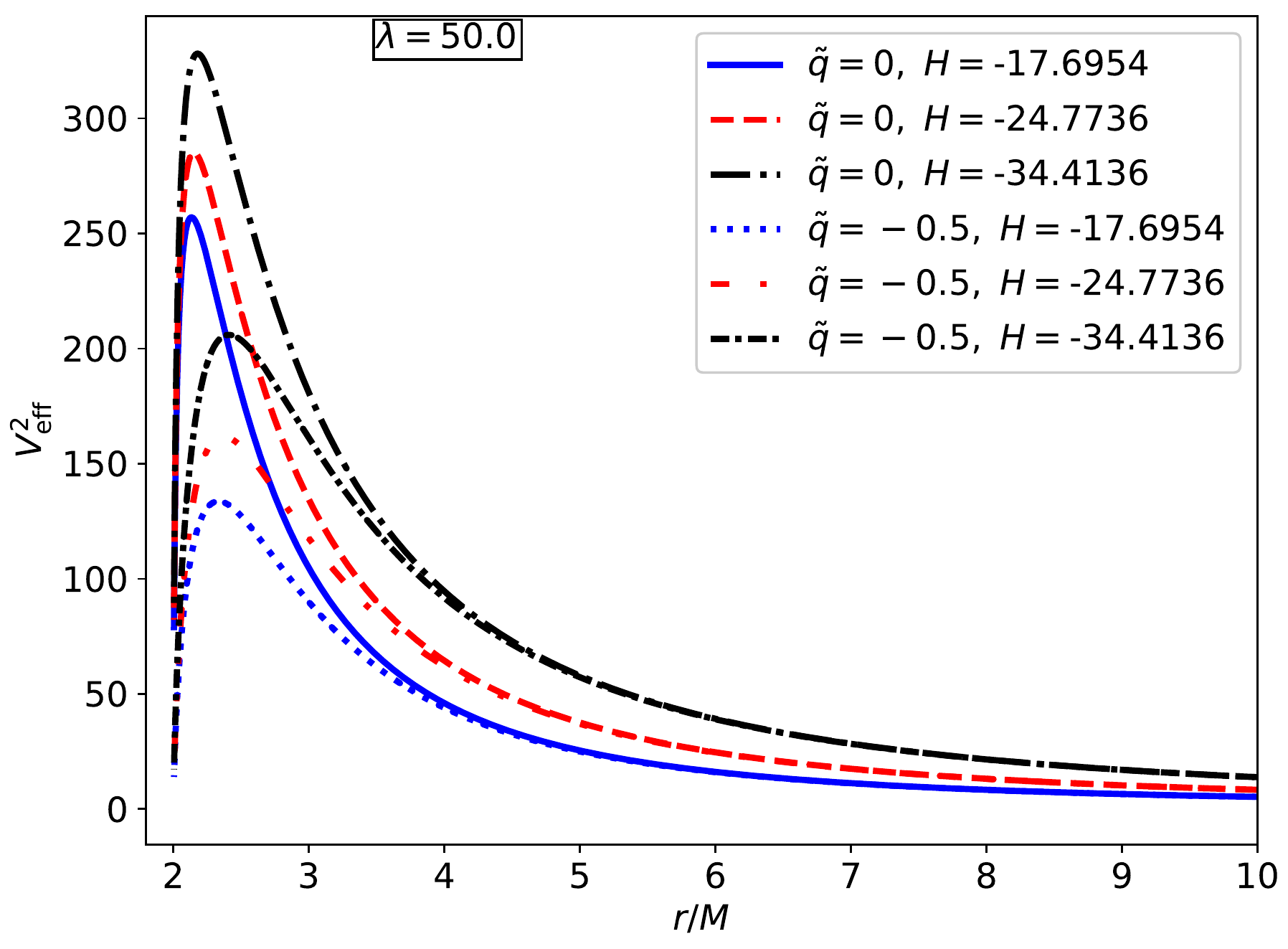}
\caption{Same as Fig.~\ref{EPDqpHn}, but here for $\tilde{q} = -0.5$. The decrease of the heights of the peaks for a negative coupling compared to the case of minimal coupling may be seen here.}
\label{EPDqnHn}
\end{figure}

For a particle having its angular momentum oriented anti-parallel to magnetic dipole, the effective potential has only one maximum (very close to the horizon) and a minimum far away from the horizon. Hence, such a particle will effectively not see any potential well. If the energy $\tilde{E}$ of the particle is greater than the maximum, it will plunge into the black hole, while if it is smaller, it will be scattered away. There also exists a possibility of an unstable circular orbit very near the event horizon (if the particle's energy is equal to the maximum), and a stable circular orbit far away from the event horizon (for less energetic particles). The effects of nonzero $\tilde{q}$ on the scattering of particles are similar to what has been discussed for the case $H>0$ (see Figs.~\ref{EPDqpHn} and \ref{EPDqnHn}).

\subsection{Uniform Magnetic Field} \label{sec:UMF}

\begin{figure}[t]
\centering
\includegraphics[scale=0.45]{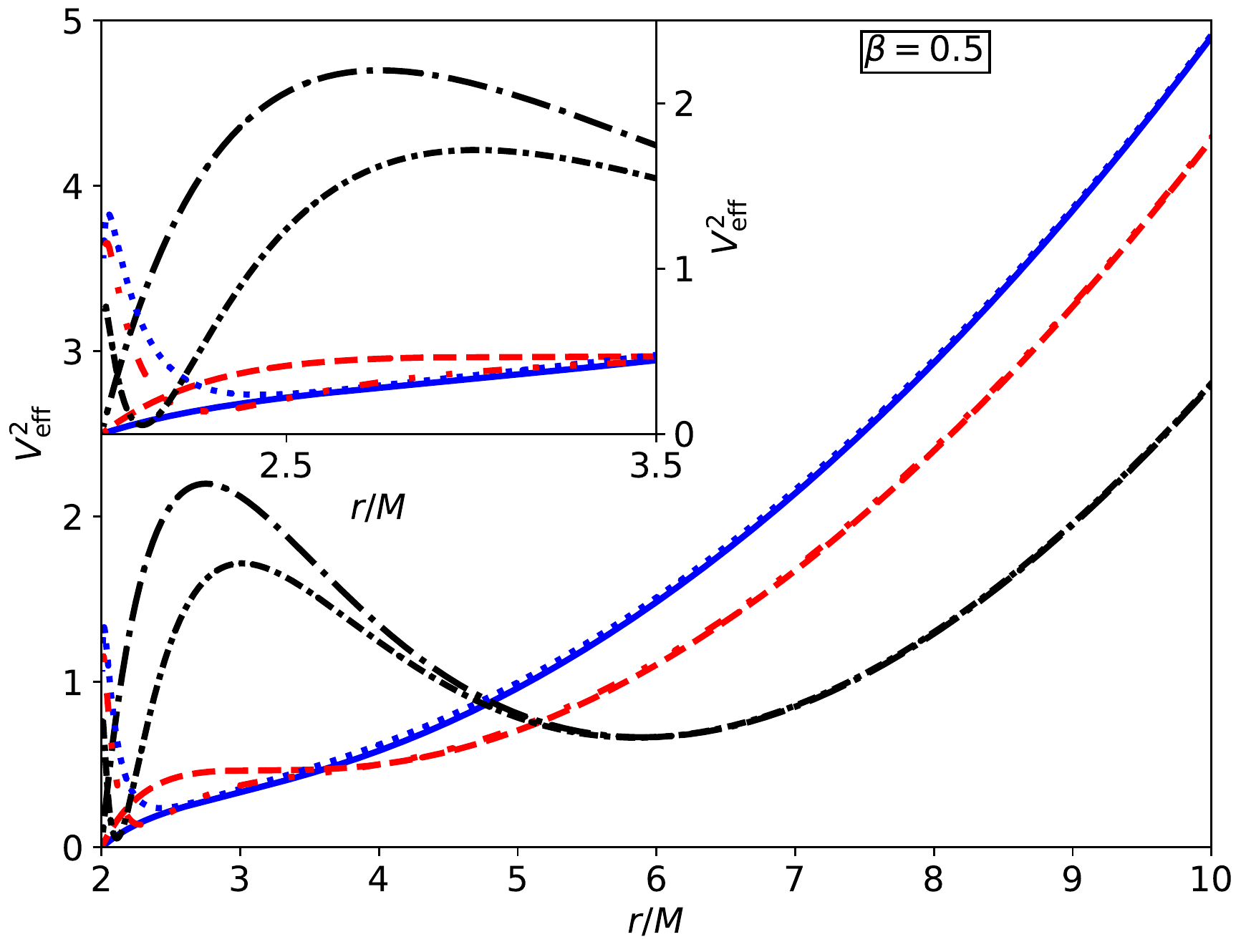}
\caption{The effective potential for the motion of a charged particle in the equatorial plane of the Schwarzschild spacetime in an asymptotically uniform magnetic field for the magnetic field parameter $\beta=0.5$, for the case of the magnetic field and the angular momentum being parallel. The colors blue, red and black corresponds to the angular momentum parameter ($H$) values $2.36$, $4.14$ and $9.2$, respectively. The {\it solid, dashed, dashdotted} curves are for $\tilde{q}=0.0$ and {\it dotted, dashdotdotted, densely dashdotted} curves are for $\tilde{q}=0.5$, respectively. Because of a lack of space in this figure we have presented these line styles in the {\it legend} of Fig.~\ref{EPUqnHp} (the only difference being the sign $\tilde{q}$, while its magnitude is the same). To highlight the two interesting features (one extra maximum and one additional minimum near the horizon) in the effective potential due to positive coupling, the inset plot is included. Only the right-side labeling of the vertical axis corresponds to the inset plots.}
\label{EPUqpHp}
\end{figure}

\begin{figure}[t]
\centering
\includegraphics[scale=0.45]{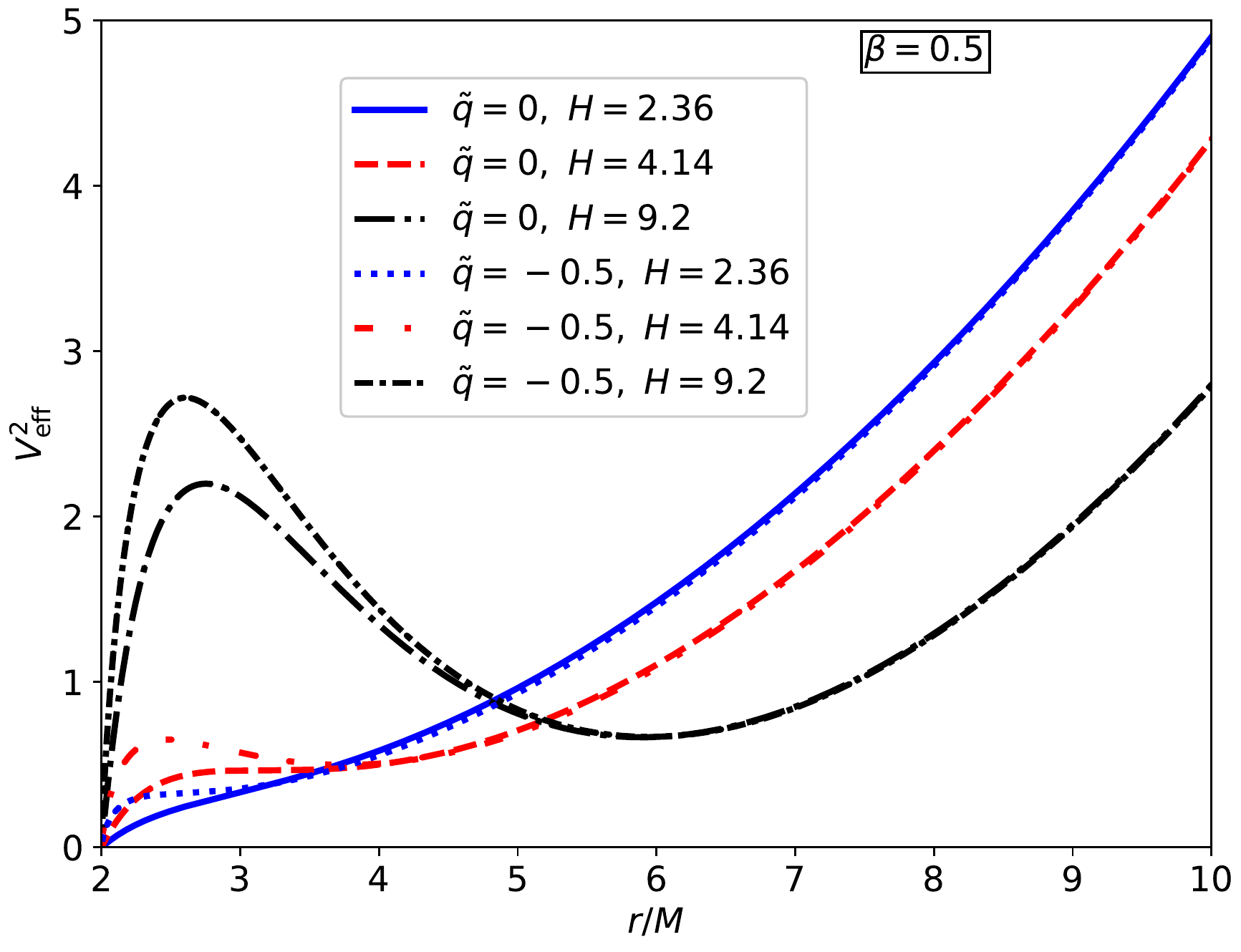}
\caption{Same as Fig.~\ref{EPUqpHp}, but now for $\tilde{q}=-0.5$. 
Unlike for the case of positive $\tilde{q}$, here for $\tilde{q}=-0.5$ the effective potentials do not differ much from the case of the minimal-coupling scenario.}
\label{EPUqnHp}
\end{figure}

Again, we point out the existence of an analytical expression for $A_{\phi}(r, \theta)$ for the case of the minimal-coupling scenario \cite{Frolov:2010mi}, 
\begin{equation}
A_{\phi} = \frac{B_{0}}{2}r^{2}\sin^{2}\theta\,,
\label{fourpotUn}
\end{equation}
where $B_{0}$ is the strength of magnetic field at infinity. To obtain the effective potential numerically for the case of non-minimal coupling, we follow the same set of procedures as described earlier for the case of a dipole field. We define one more dimensionless quantity, $\beta\equiv qB_{0}M/m$, and will make use of $H$ as defined earlier. For the radial distance we will use a dimensionless parameter, $\rho=r/M$ , for which $\rho_{+}$ and $\rho_{-}$ will be used to denote the radius of the {\it isco} when the angular momentum is parallel and anti-parallel to the magnetic field, respectively. Unlike in the case of a dipole magnetic field, the uniform magnetic field at large distances from the horizon plays an important role as a barrier for the particles coming from infinity (see Figs.~\ref{EPUqpHp} and \ref{EPUqnHp}).

\begin{figure}[t]
\centering
\includegraphics[scale=0.45]{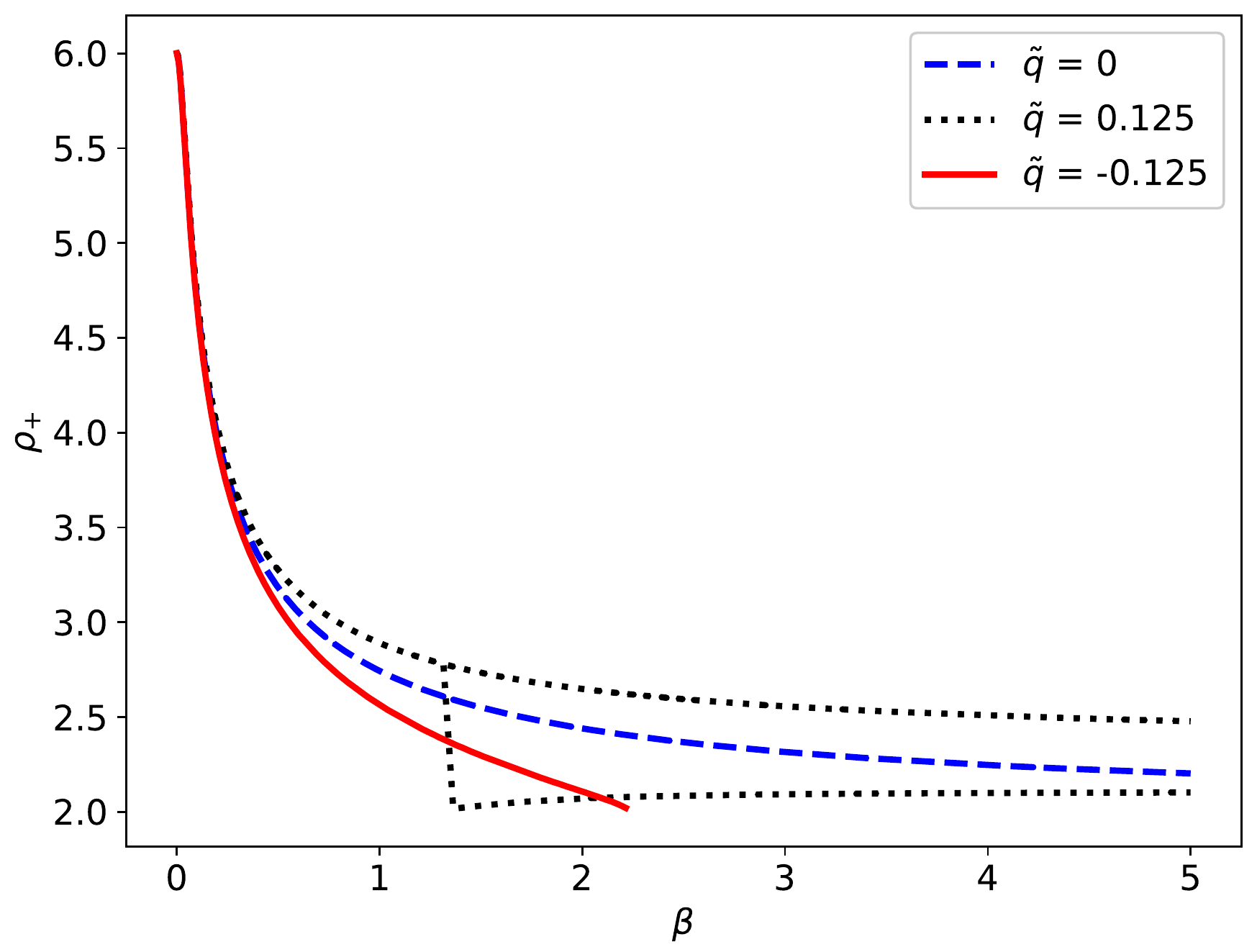}
\caption{The {\it isco} radii($\rho_{+}$) as a function of $\beta$ when the angular momentum of the particle is parallel to the uniform magnetic field (asymptotically). The colors black, blue and red are for the values of $\tilde{q}=$0.125,\,0.0 and -0.125, respectively. As expected in all three (different values of $\tilde{q}$) cases, the {\it isco} radius should be $6M$ when there is no magnetic field (i.e.~when $\beta$=0). For the case of minimal coupling with an increase in the strength of the magnetic field parameter $\beta$, the {\it isco} radius saturates for $r_{ isco} \rightarrow 2M$ to $\beta \rightarrow \infty$.}
\label{iscoUni1left}
\end{figure}

\begin{figure}[t]
\centering
\includegraphics[scale=0.45]{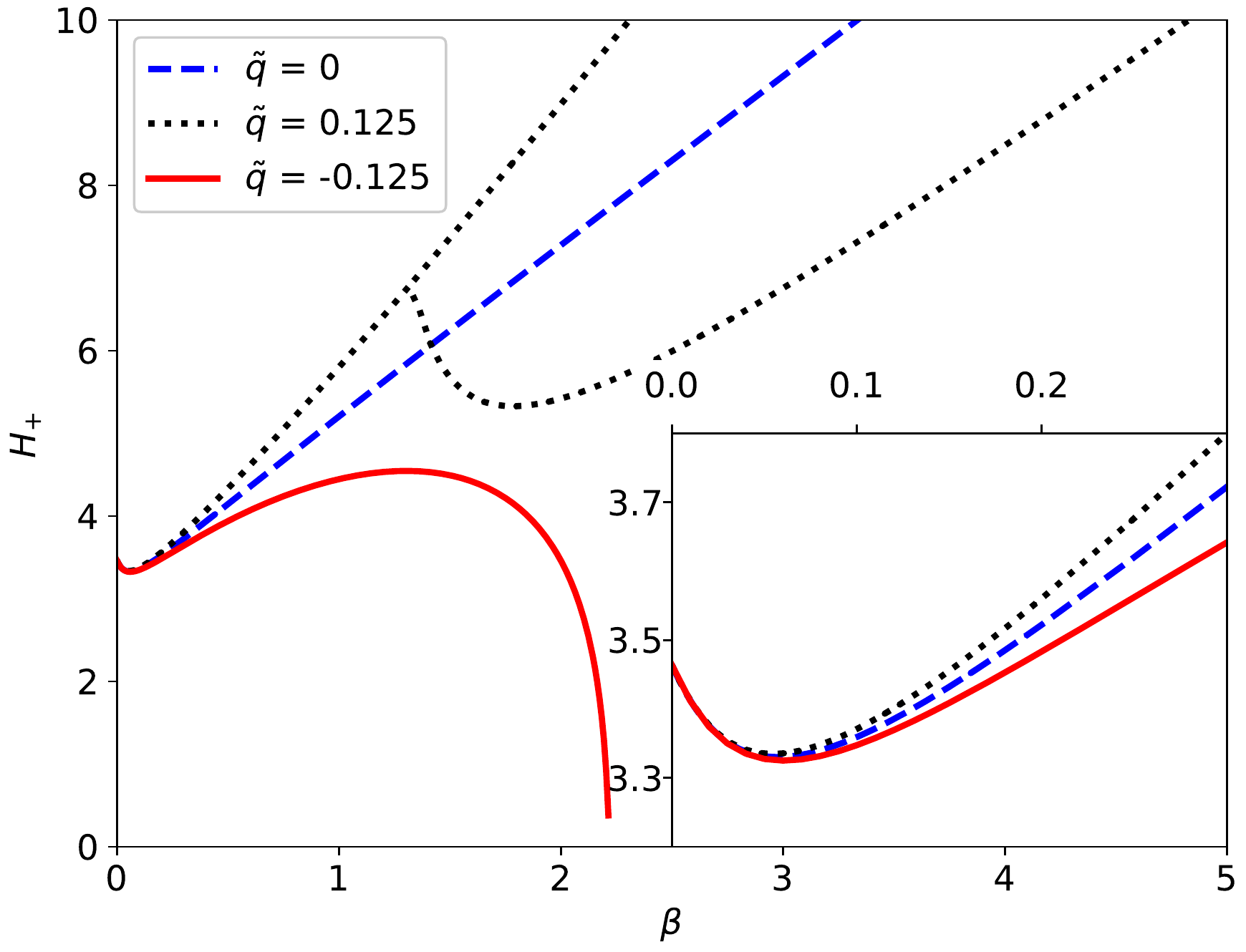}
\caption{The angular momentum parameter ($H_{+}$) for {\it isco} as a function of $\beta$. The colors black, blue and red are for the values of $\tilde{q}=0.125$, $0.0$ and $-0.125$, respectively. As expected, in absence of any magnetic field the angular momentum here is $+\sqrt{12}M$.}
\label{iscoUni2left}
\end{figure}

\begin{figure}[t]
\centering
\includegraphics[width=0.5\columnwidth]{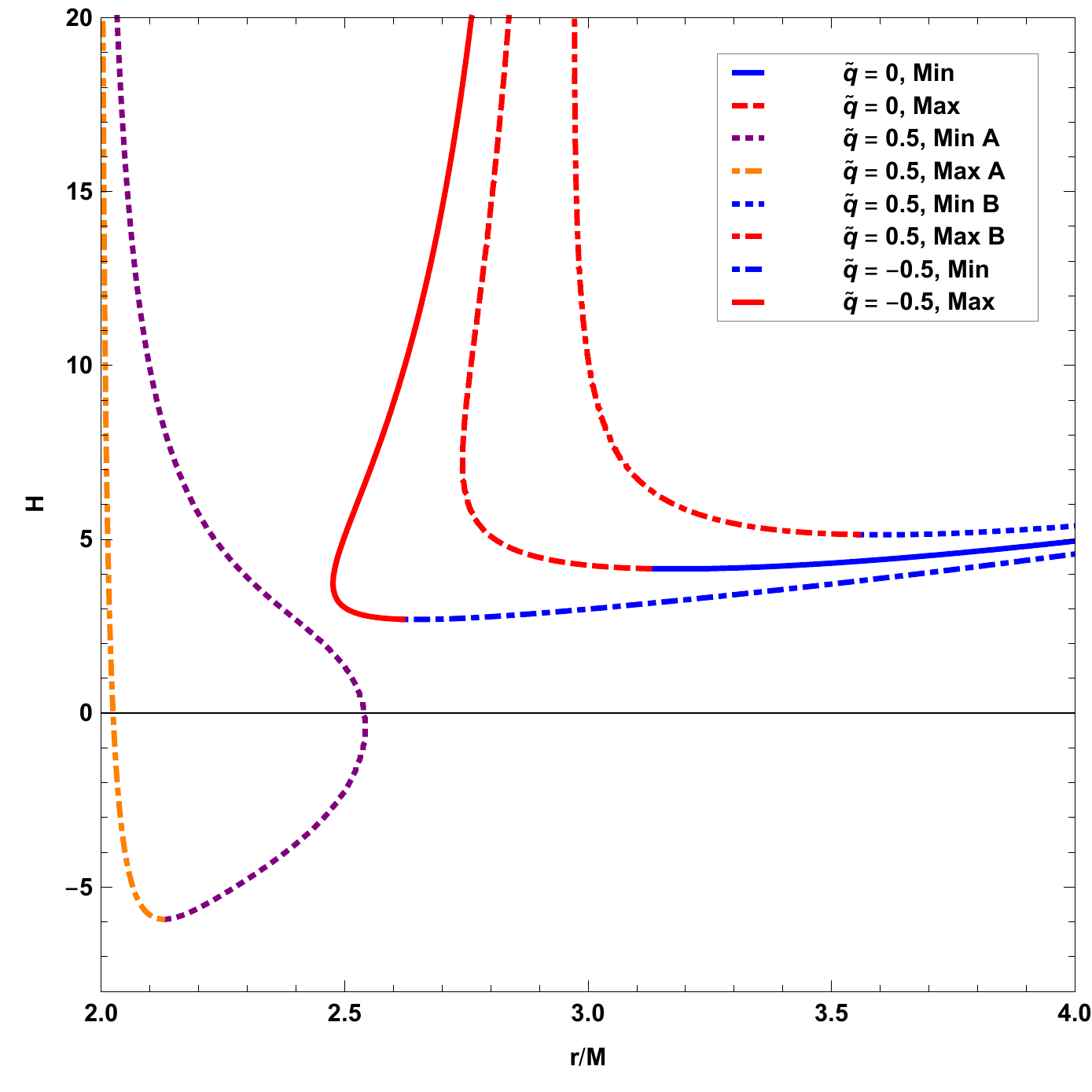}
\caption{Location of the extrema of the effective potential $V_{\rm eff}^{2}$ for $\beta=0.5$ in the $r$-$H$ parameter space (upper branch) for different values of $\tilde{q}$. For $\tilde{q}=0.5$ two minima/maxima (denoted as Min A/B and Max A/B, respectively) exist. Note that while this upper branch has strictly $H>0$ in the case of non-minimal coupling, for $\tilde{q}=0.5$ it obtains values with $H<0$.}
\label{fig:minmaxplot_beta0p5}
\end{figure}

\subsubsection{Case 1: The Angular Momentum is Parallel to the Magnetic Field}

For our investigation of non-minimal coupling of an asymptotically uniform magnetic field with Schwarzschild spacetime, we have used the values of parameters $\beta$ and $H$ from the caption of Fig.~3 from \cite{Frolov:2010mi}. We also note that the quoted $l \simeq 3.22$ in that figure caption is a typographical error and we find it to be $l \simeq 4.6$. The parameters $\rho$, $\beta$ and $H$ defined here corresponds to $2\rho$, $b$ and $2l$ in \cite{Frolov:2010mi}, respectively. We will focus our discussion mainly on the case for $\tilde{q} \neq 0$, as the $\tilde{q} = 0$ case is nicely presented in \cite{Frolov:2010mi}.

In the case of a minimally coupled magnetic field, for lower values of $H$ there are, at first, no extrema of the effective potential present, while for an increase in $H$ we see one extremum (inflection point), and increasing $H$ even further results in one maximum near and one minimum away from the horizon (as can be seen in Fig.~\ref{EPUqpHp} for $H \simeq 2.36,\,4.14,\,9.2$). For a further increase in $H$, both of these extrema shift outwards from the horizon with the increased values of the effective potential. A positive $\tilde{q}$ introduces none, one or two more extrema near the horizon, depending on $H$. So, for those values of $\beta$ and $H$, for which there are no extrema at all in the case of minimal coupling, a positive $\tilde{q}$ can introduce an inflection point or a minimum and hence lead to the existence of an {\it isco} or a potential well. For those values of $\beta$ and $H$, for which there is an inflection point, a positive $\tilde{q}$ can introduce a minimum and hence a potential well. Finally, for those higher values of $\beta$ and $H$, for which there is a maximum and a minimum, a positive $\tilde{q}$ can introduce one more maximum and one more minimum and hence one more potential well. Effectively, a positive $\tilde{q}$ introduces bound gyrating trajectories (and stable/unstable circular orbits) very near the horizon, separated from the same type of trajectories away from the horizon. This observation can have its implications in modeling of accretion disks and while studying any emission (possibly synchrotron emission, see below) from particles gyrating in the bounded orbits.

A negative $\tilde{q}$ introduces at most one inflection point or minimum near the horizon if no such features are there in case of minimal coupling. For an already existing potential well, a negative $\tilde{q}$ increases the depth (see Fig.~\ref{EPUqnHp}). Unlike the case of $\tilde{q}>0$, here we do not see two sets of bounded orbits, but with an increased depth of the potential well, trapped particle will have higher energies compared to the case of minimal coupling.

In Fig.~\ref{iscoUni1left} we have plotted the dependence of the radius of the {\it isco} on the magnetic field parameter $\beta$. In the figure we clearly can see the {\it isco} radius at $6M$ for zero magnetic field, then, for an increasing magnetic field, it shifts towards the horizon. As expected, for a positive $\tilde{q}$ and in the case of smaller values of magnetic field (i.e.~values up to which the non-minimal coupling does not introduce any inner minima, but rather just shifts the extrema of the minimal coupling scenario outwards, see Fig.~\ref{EPUqpHp}) the {\it isco} radii are larger than those for the minimal-coupling case. Above a certain strength of the magnetic field, the non-minimal coupling starts introducing additional extrema near the horizon (this aspect can be seen in Fig.~\ref{iscoUni1left} around $\beta \simeq 1.3$. We choose $|\tilde{q}|=0.125$ instead of $0.5$ to show this transitory feature clearly, since for $\tilde{q} = 0.5$ this feature shows near $\beta\sim0.1$). Therefore, we see that beyond certain values of $\beta$ there are two sets of stable circular orbits (well within $6M$ radius) separated from each other. The termination of plot for negative $\tilde{q}$ should be interpreted as {\it isco} grazing the horizon and being occupied with angular momentum of the particles $H \rightarrow 0$. Because of limited resolution ($\Delta H$) used in our code, this curve is terminating there. This feature is not surprising, as we carefully observe the plot of the effective potential in the Fig.~\ref{EPUqnHp} for $H=2.36$. Like in the case of minimal coupling for $\beta \rightarrow \infty$ \cite{Frolov:2010mi}, we expect the radius of {\it isco} to saturate at some value (depending on $\tilde{q}$) near $2M$ (the Schwarzschild radius).

In Fig.~\ref{iscoUni2left} we have plotted the angular momentum for {\it isco} as a function of magnetic field parameter $\beta$. We can see that, starting from the angular momentum of $\sqrt{12}M$ for $\beta=0$, the particles with lower angular momentum initially occupy the {\it isco}, while with the increase in the strength of the magnetic field particles of higher angular momentum orbit in the {\it isco} as well. Again, the transitory feature near $\beta \simeq 1.3$ for $\tilde{q} = 0.125$ reflects the introduction of an inflection point or a shallow potential well near the horizon because of non-minimal coupling (see the inset of Fig.~\ref{EPUqpHp}). The termination of the curve for $\tilde{q} = -0.125$ near $\beta \simeq 2.2$ (with $H \simeq 0.4$) reflects the fact that for the continuation of this curve one needs a progressively finer and finer resolution ($\Delta H$) requiring heavy computation. As the gained information is not justifying such heavy computations, we have terminated our calculations beyond these limits.

As a concluding remark and summary to this section we present the change of the position of the minima and maxima of the effective potential in the $r$-$H$ phase space in Fig.~\ref{fig:minmaxplot_beta0p5}. As already described, the introduction of $\tilde{q} \neq 0$ indeed greatly modifies the form of the effective potential, in particular by creating additional extrema and also shifting this ``upper branch" (i.e.~the branch of $H>0$ for $\tilde{q}=0$) to negative values of $H$. In fact, this modification can alter observational signatures which, in turn, may be used in the future to derive limits on $\tilde{q}$, as described in Sec.~\ref{sec:SynchRad}.

\subsubsection{Case 2: The Angular Momentum is Anti-Parallel to the Magnetic Field}

\begin{figure}[t]
\centering
\includegraphics[scale=0.45]{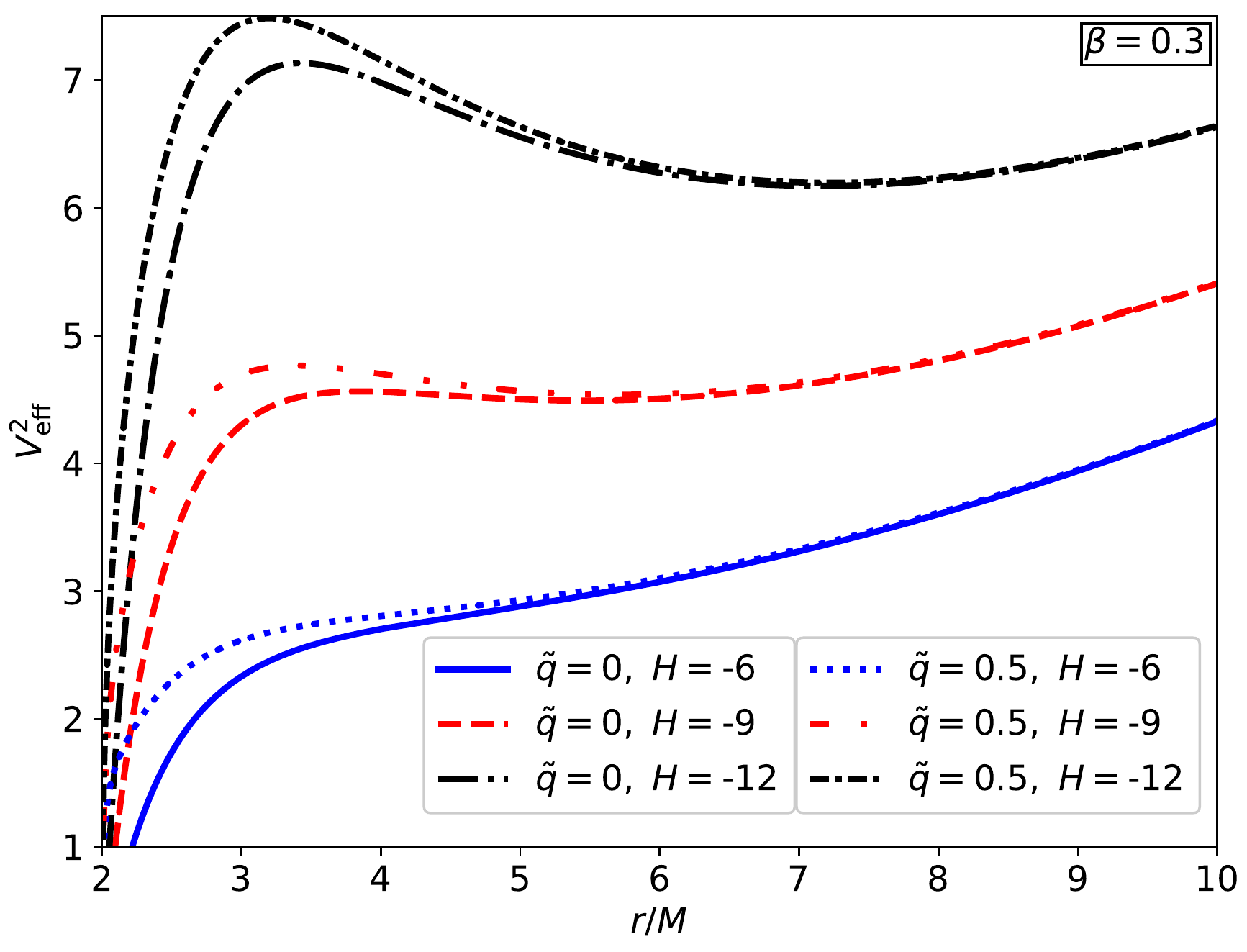}
\caption{The effective potential for the motion of a charged particle in the equatorial plane of the Schwarzschild spacetime in an asymptotically uniform magnetic field with magnetic field parameter $\beta = 0.3$, for the case of the magnetic field being anti-parallel to angular momentum.}
\label{EPUqpHn}
\end{figure}

\begin{figure}[t]
\centering
\includegraphics[scale=0.45]{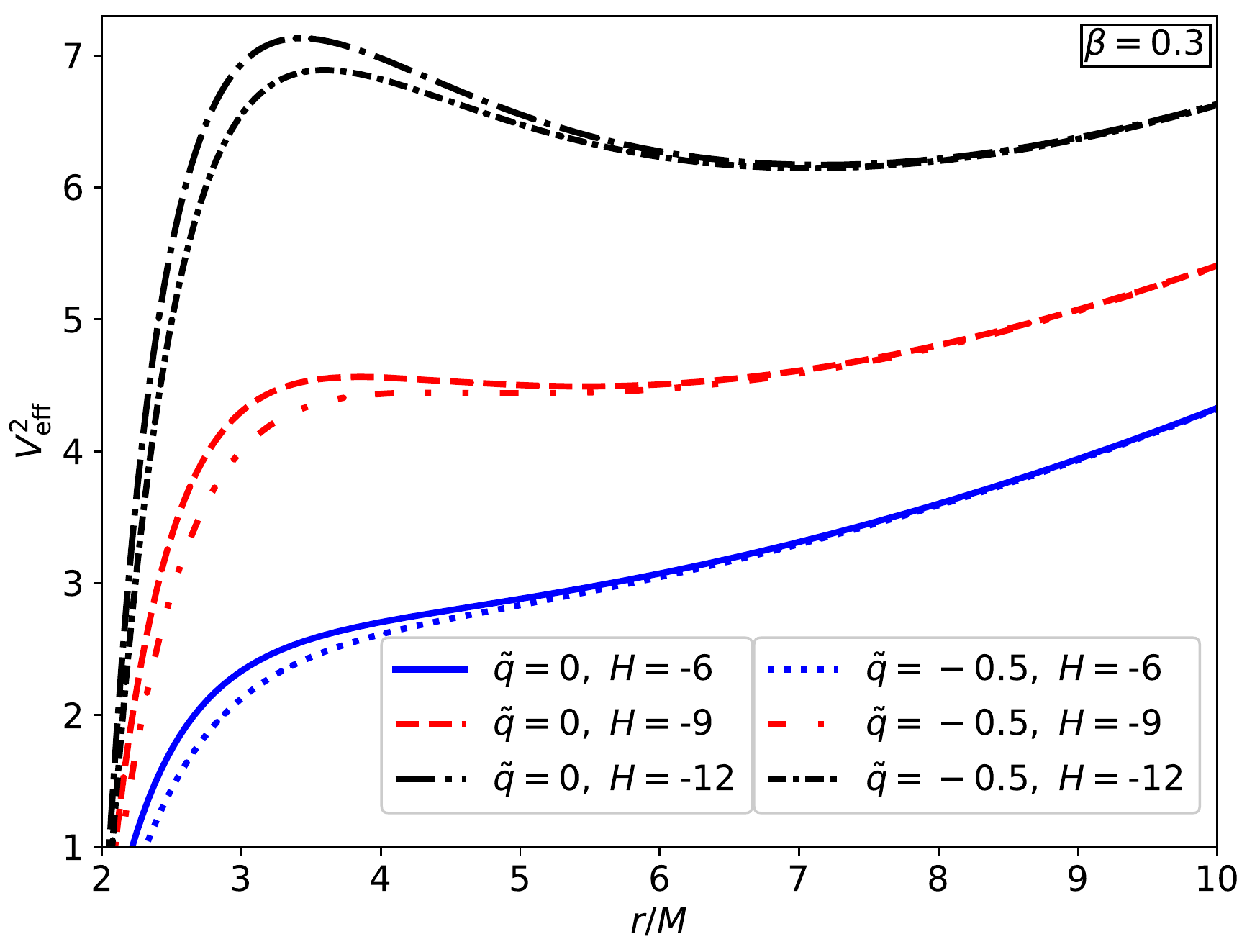}
\caption{Same as Fig.~\ref{EPUqpHn}, but now for $\tilde{q}=-0.5$.}
\label{EPUqnHn}
\end{figure}

\begin{figure}[t]
\centering
\includegraphics[scale=0.45]{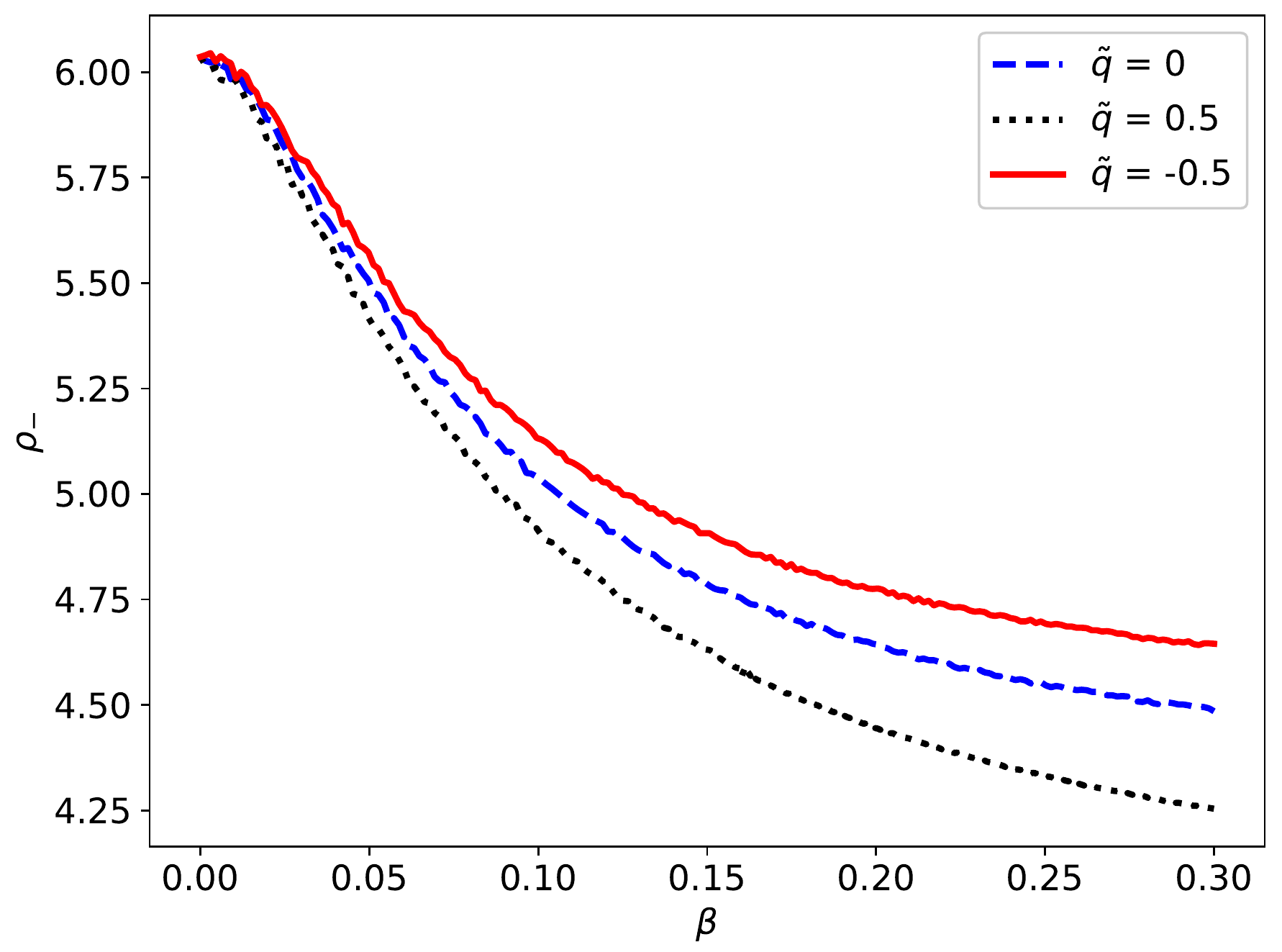}
\caption{The {\it isco} radii ($\rho_{-}$) as a function of $\beta$ when angular momentum of the particle is anti-parallel to the uniform magnetic field (asymptotically). The colors black, blue and red are for the values of $\tilde{q}=0.5$, $0.0$ and $-0.5$, respectively. As expected, in the absence of a magnetic field the radius should be $6M$. For the case of minimal coupling, for an increase of the strength of the magnetic field parameter $\beta$, the {\it isco} radius saturates as $r_{ isco} \rightarrow 4.3M$ for $\beta \rightarrow \infty$ \cite{Frolov:2010mi}.}
\label{iscoUni1right}
\end{figure}

\begin{figure}[t]
\centering
\includegraphics[width=0.5\columnwidth]{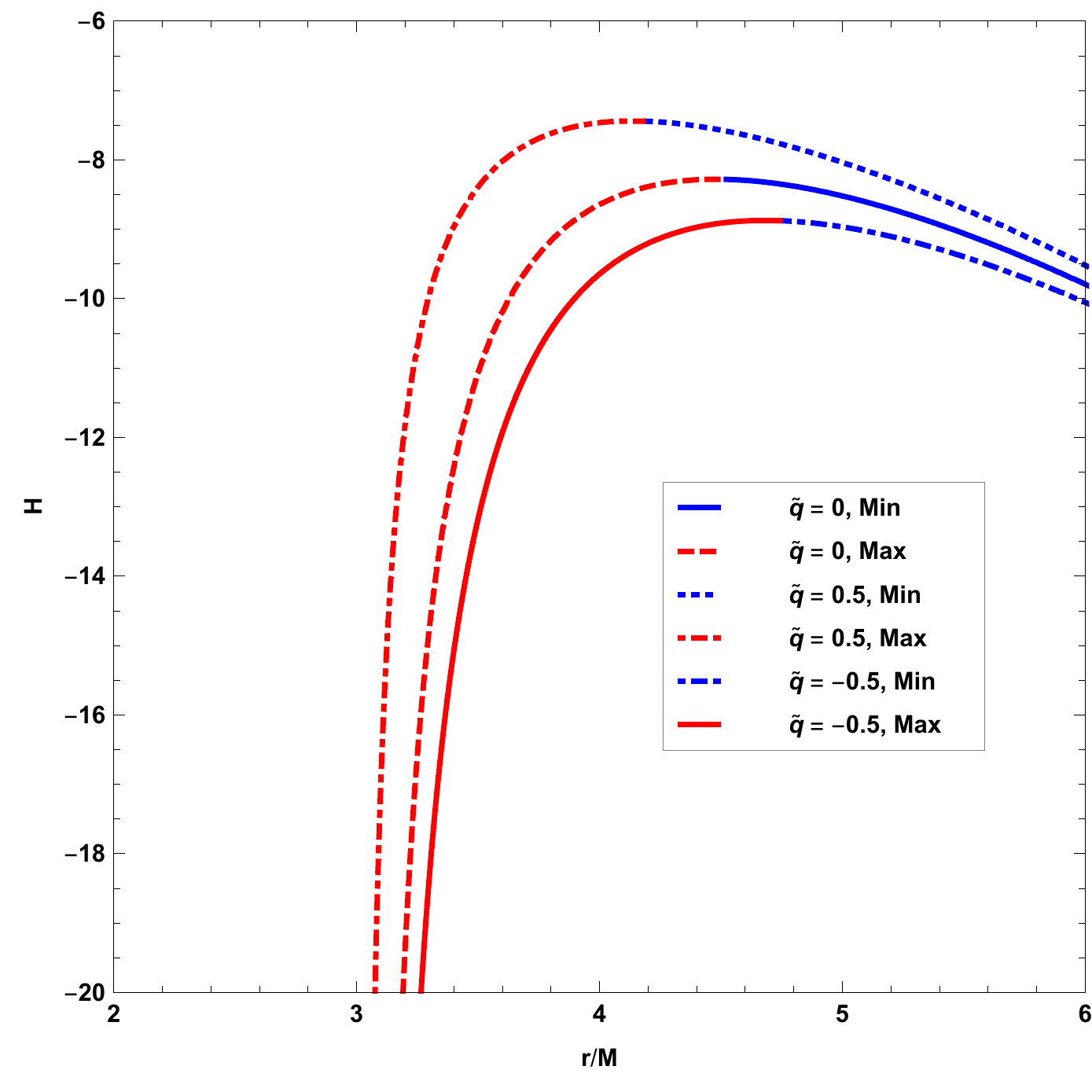}
\caption{Location of the extrema of the effective potential $V_{\rm eff}^{2}$ for $\beta=0.3$ in the $r$-$H$ parameter space (lower branch) for different values of $\tilde{q}$. In contrast to the upper branch (cf.~Fig.~\ref{fig:minmaxplot_beta0p5}), the change with $\tilde{q}$ of the overall structure here is much smaller.}
\label{fig:minmaxplot_beta0p3}
\end{figure}

\begin{figure}[t]
\centering
\includegraphics[scale=0.45]{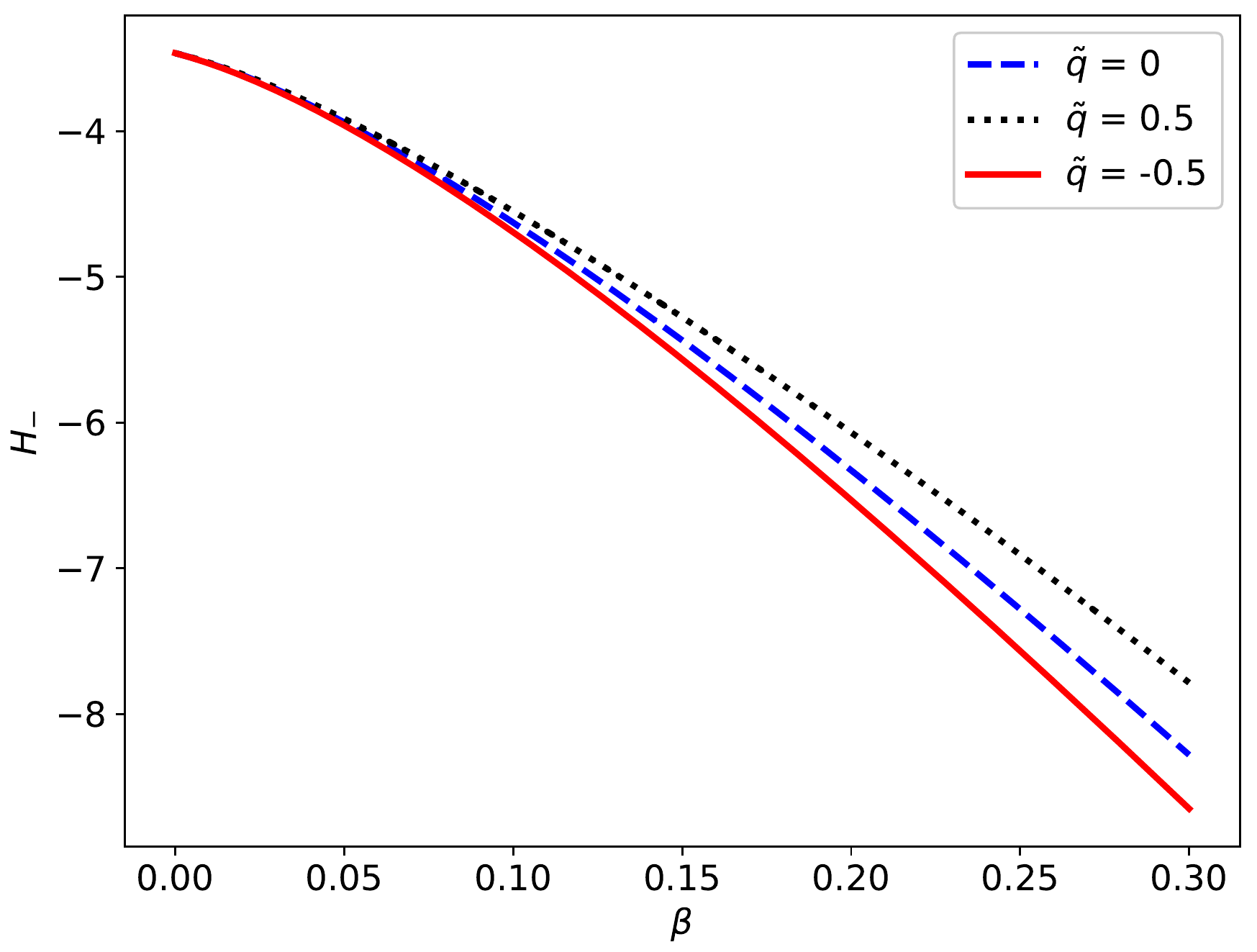}
\caption{The angular momentum parameter ($H_{-}$) for the {\it isco} as a function of $\beta$. The colors black, blue and red are for the values of $\tilde{q}=0.5$, $0.0$ and $-0.5$, respectively. As expected, in the absence of a magnetic field the angular momentum is $-\sqrt{12}M$.}
\label{iscoUni2right}
\end{figure}

When the angular momentum of a particle is anti-parallel to the magnetic field we hardly see any interesting features in the effective potential in both coupling scenarios. Results in the case of non-minimal coupling are just slightly enhanced/suppressed quantitatively. We can speculate about the possible reason for such a situation to be the saturation of the {\it isco} radius ($\rho_{-}$) comparatively (with the case of parallel angular momentum) far from the horizon, $\rho_{-}\rightarrow 4.3M$ as $\beta\rightarrow \infty$ \cite{Frolov:2010mi}.

In Figs.~\ref{EPUqpHn} and \ref{EPUqnHn} we have presented a few illustrative plots for the effective potential. Depending on $\beta$ and $H$, there exist a maximum (and hence an unstable circular orbit) near the horizon, followed by a minimum (and hence a stable circular orbit and/or potential well for particle trappings) away from the horizon. A positive/negative $\tilde{q}$ raises/lowers the height of the maxima and shifts inward/outwards. The effects of positive/negative $\tilde{q}$ on the minima is to shift it inwards/outwards and hardly makes any change in its height. 

In Fig.~\ref{iscoUni1right} we have plotted the radii ($\rho_{-}$) of the {\it isco} as a function of the parameter $\beta$. A positive/negative value of the coupling constant $\tilde{q}$ allows the {\it isco} to be closer/farther to/from the event horizon when compared with the case of minimal coupling. Correspondingly, the energy of particles trapped in unstable/stable circular orbits or in gyrating bound orbits will be different for the two kinds of coupling. In these figures, for a zero magnetic field ($\beta=0$) we again can see the {\it isco} radius $6M$ and the corresponding angular momentum $-\sqrt{12}\,M$.

Finally, in a similar way as for the previous section, in Fig.~\ref{fig:minmaxplot_beta0p3} we present the behavior of the extrema of the effective potential in the $r$-$H$ parameter space (for the lower branch, i.e.~$H<0$ with respect to $\tilde{q}$. While a modification is clearly visible, in contrast to the upper branch described before the basic shape remains the same and no additional extrema are introduced.

\subsubsection{Energy of a Charged Particle in a Marginally Stable Circular Orbit}

\begin{figure}[t]
\centering
\includegraphics[scale=0.45]{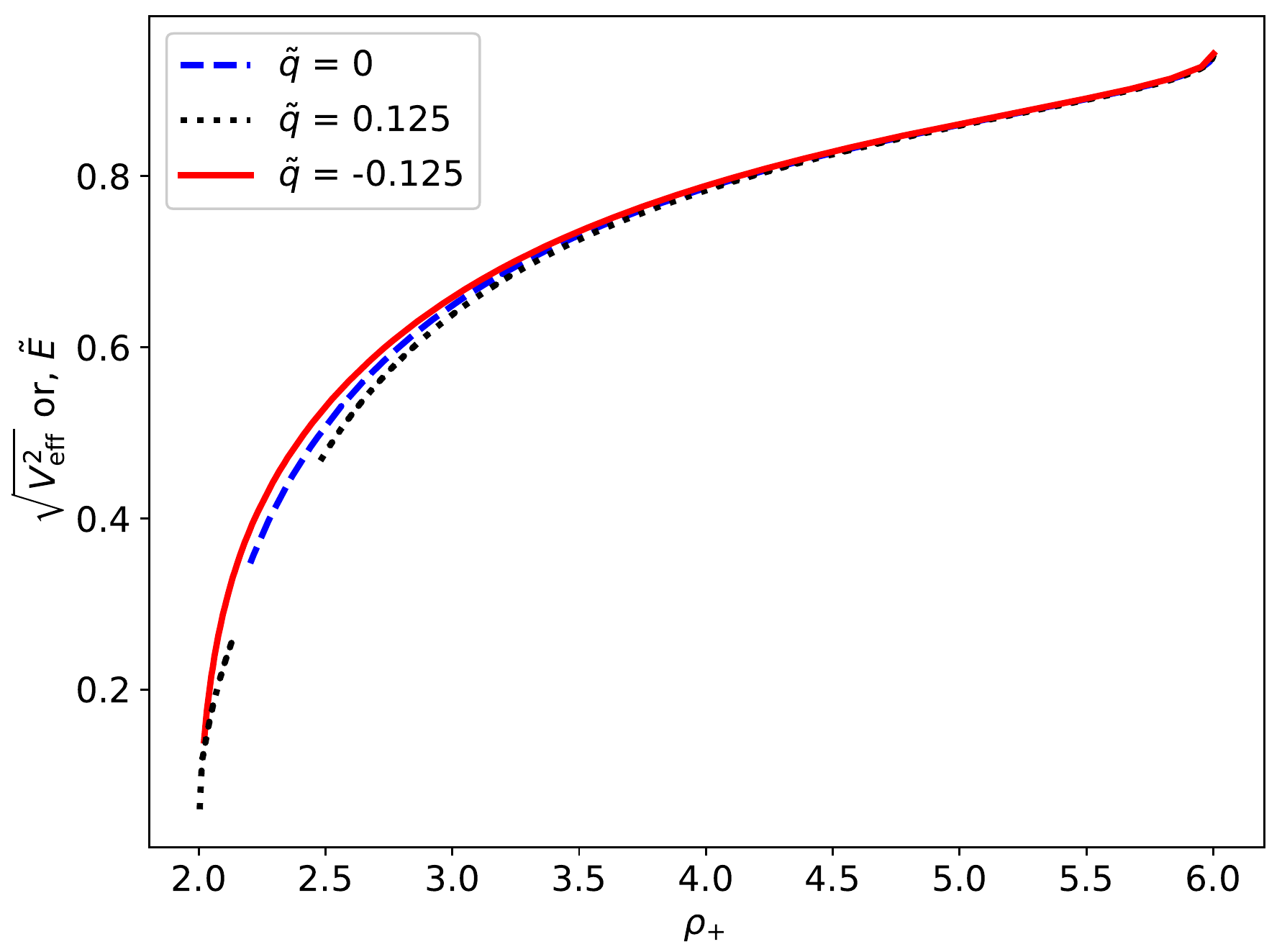}
\caption{Energy of a charged particle in an {\it isco} as function of the {\it isco} radius when its angular momentum is parallel to the uniform magnetic field.}
\label{iscoUniEngParal}
\end{figure}

As a prequel to the study of the collision of charged particles, in this section we briefly discuss the energetics of charged particles moving in a marginally stable circular orbit. In the case of the angular momentum being parallel to the asymptotically uniform magnetic field, a non-minimal coupling makes it possible to have the {\it isco} closer to the horizon compared to the case of minimal coupling (of course above a certain value of magnetic field strength this feature is common to both the positive and the negative charge, see Figs.~\ref{iscoUni1left} and Fig.~\ref{iscoUniEngParal}). The break/gap in the energy plot for $\tilde{q} = 0.125$ is understandable if we observe Fig.~\ref{iscoUni1left}, since for $\beta < 5$ there are no {\it isco} radii in the range $\sim 2.12-2.47$, while with the increase in $\beta$ this gap will fill.

\begin{figure}[t]
\centering
\includegraphics[scale=0.45]{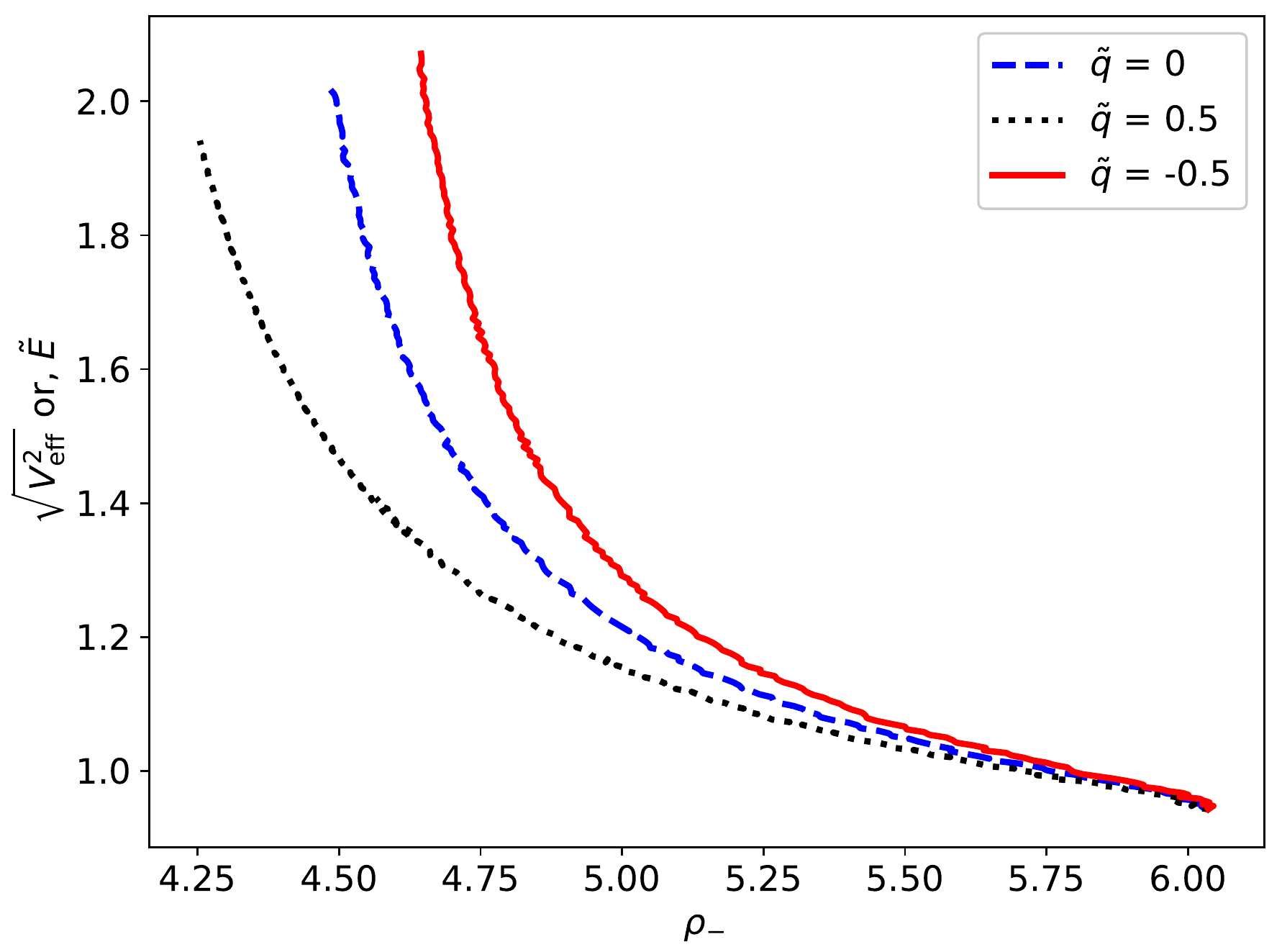}
\caption{Energy of a charged particle in an {\it isco} as function of the {\it isco} radius, when its angular momentum is anti-parallel to the uniform magnetic field.}
\label{iscoUniEngAntiParal}
\end{figure}
When the angular momentum is anti-parallel to the magnetic field, similar to the results in the case of minimal coupling, the non-minimal coupling also enables trapping of the particles with higher energies ($\tilde{E} > 1$, see Fig.~\ref{iscoUniEngAntiParal}). For a positive/negative value of the coupling constant $\tilde{q}$ and any given $\beta$ these high energy particles can move in closer/farther {\it isco}s in comparison to the case of minimal coupling.

\section{Particle Collision/Acceleration} \label{sec:PartCollAcc}

\begin{figure}[t]
\centering
\includegraphics[scale=0.45]{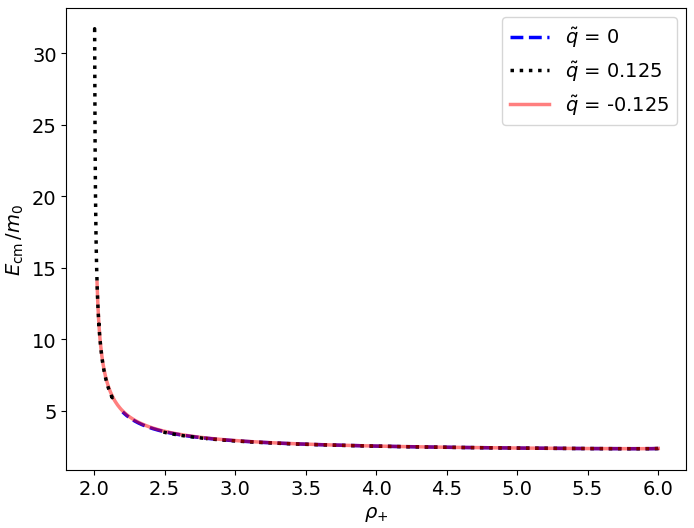}
\caption{The center of mass energy for a head-on collision of two charged particles of equal mass $m_{0}$, equal and opposite charge, moving in opposite directions, therefore having equal angular momenta (when the angular momenta of these charged particles are parallel to the magnetic field) as a function of the {\it isco} radii.}
\label{iscoUniColParal}
\end{figure}

Taking the case of an asymptotically uniform magnetic field, we consider the collision of two charged particles of equal mass ($m_{0}$), opposite charge and moving in opposite directions on an {\it isco} (therefore having the same angular momentum). For this we calculate the center of mass energy of the two colliding particles, given by 
\begin{equation}
E_{\rm cm}^{2} = 2m_{0}^{2}\left(1 - g_{\mu\nu}u^{\mu}_{1}u^{\nu}_{2}\right)\,,
\end{equation}
where $u^{\mu}_{1}$ and $u^{\mu}_{2}$ are their respective 4-velocities. For this configuration of the collision the 4-velocities of the particles take the form
\begin{equation}
u^{\alpha}_{\pm} = \left(-\frac{{\rm d}t}{{\rm d}\tau},\,0,\,0,\,\pm\frac{{\rm d}\phi}{{\rm d}\tau} \right)
\end{equation}
and are subject to the normalization $u^{\alpha} u_{\alpha} = -1$. Using Eqs.~(\ref{eng}), (\ref{amom}) and (\ref{Veff}), we get
\begin{equation}
\frac{E_{\rm cm}}{m_{0}} = \sqrt{2}\left(1 - \frac{2M}{r}\right)^{-1/2}\sqrt{\tilde{E}^{2} + V^{2}_{\rm eff}}\,.
\label{Ecm1}
\end{equation}
In Fig.~\ref{iscoUniColParal} we have plotted the center of mass energy over the rest energy of a charged particle as function of the {\it isco} radius when the angular momentum of the particles is parallel to the magnetic field. For the case of minimal coupling we do not see much gain in energy \cite{Frolov:2011ea}. In the case of non-minimal coupling, beyond some values of magnetic field strength, near the horizon, the ratio $E_{\rm cm}/m_{0}$ dominates when compared to the case of minimal coupling. The energy of a charged particle in an {\it isco} near the horizon in case of a nonzero $\tilde{q}$ is not much larger comparatively (see Fig.~\ref{iscoUniEngParal}). However, because of the relative proximity to the horizon (when compared to the case of minimal coupling), the term $\sqrt{1 - 2M/r}$ in Eq.~(\ref{Ecm1}) enhances this ratio 
$E_{\rm cm}/m_{0}$.

\begin{figure}[t]
\centering
\includegraphics[scale=0.45]{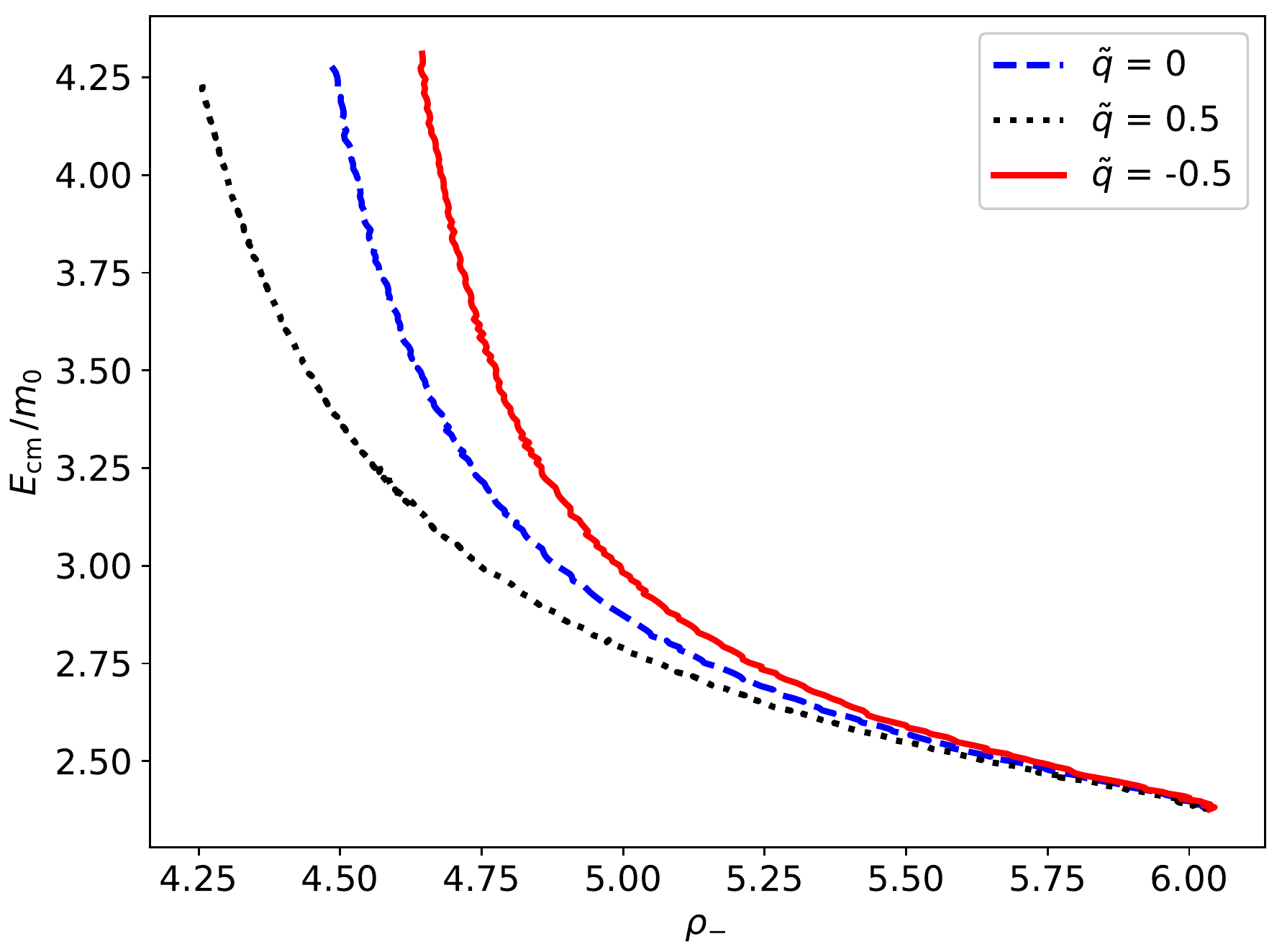}
\caption{The center of mass energy for a head-on collision of two charged particles of equal mass $m_{0}$, equal and opposite charge, moving in opposite directions, therefore having equal angular momenta (when the angular momenta of these charged particles are anti-parallel to the magnetic field) as a function of the {\it isco} radius.}
\label{iscoUniColAntiParal}
\end{figure}

The center of mass energy over the rest energy of a charged particle as function of the {\it isco} radius when the angular momentum of the particles is anti-parallel to the magnetic field is shown in Fig.~\ref{iscoUniColAntiParal}. Unlike the case of parallel angular momentum here the {\it isco} radii are away from the horizon, the effect of which can be seen if we observe the numbers on vertical axis of Fig.~\ref{iscoUniColAntiParal}. The seemingly blowing-off of $E_{\rm cm}/m_{0}$ near the minimum {\it isco} radius is not to be interpreted as a gain of energy due to the black hole, but coming from the fact that these particles are already energetic (see Fig.~\ref{iscoUniEngAntiParal}). In other words, they have high energies when measured by an observer at infinity, i.e.~the black hole is not functioning as an accelerator here \cite{Frolov:2011ea}.

\section{Possible Applications in Astrophysical and Cosmological Scenarios} \label{sec:Applications}

Due to the required amount of work and space, it is not possible to present any detailed numerical and quantitative predictions regarding the systems in which the non-minimal coupling could have its astrophysical application. While leaving such more detailed and extensive investigations for the future, in this section we will explore several possible implications of the non-minimal coupling qualitatively. Recently the non-minimal coupling of photon to the Weyl tensor was considered in \cite{Jana:2021lqe}. The observational constraint on coupling parameter thus derived has been estimated to be $\sim 0.563$. Though in this treatment a different metric was used for the description of black hole, and not the Schwarzschild metric, their estimate is not too different from the order of $\tilde{q}\sim0.5-1$, used in our work for example. Whereas in the context of the Sun and pulsar(of mass $1.33M_{\odot}$), for observable effects of non-minimal coupling the constraint on $\tilde{q}$ obtained in \cite{Prasanna:2003ix} are $\sim 3.3\times10^{9}$ and $\sim1.5\times10^{2}$. Clearly for the later case we can not claim the validity of our present work  Applications of this effect for objects like a neutron star, the sun and Earth are discussed in \cite{Pavlovic:2018idi}, such that we will not discussion them here. In \cite{Pavlovic:2018idi} the application to primordial black holes is also discussed, but nevertheless here we would like to present it in a different light.

One of the central issues which appears when discussing the potential observational effects of the non-minimal coupling is the value of the coupling parameter. Broadly speaking, one approach to this question is to consider the coupling constant to be a free parameter that has to be obtained or constrained phenomenologically, i.e.~from observations of astrophysical sources, as have been considered in the literature \cite{1971PhLA...37..331P, Prasanna:1973xv}. On the other hand, Drummond \& Hathrell \cite{Drummond:1979pp} have presented a derivation of the non-minimal coupling from first principles, demonstrating it to be the consequence of the QED vacuum polarization on the curved spacetime. In their derivation they have found that the coupling constant is determined to be $q_{3} = - \alpha{\lambda_{e}^{2}}/90\pi\,$ $\sim -10^{-28}\,{\rm m}^{2}$, where $\alpha$ is fine structure constant and $\lambda_{e}$ is the Compton wavelength of the electron. It is, however, important to take into account that this value is determined for a very specific and elementary setting of a photon propagating in vacuum on the curved spacetime. It is not at all simple to determine what is the connection between this very simple case of one photon and a very complex setting of photons leading to macroscopic magnetic field distributions, such as the dipole field. Therefore, it is not possible to simply claim that the coupling parameter for macroscopic configurations of magnetic fields -- i.e.~the ones encountered in astrophysics -- is related to the value calculated in \cite{Drummond:1979pp}. In principle, the value of the coupling parameter could even vary for different field strengths and configurations. It therefore seems that the value of the coupling parameter for the macroscopic magnetic field configurations of interest is basically unknown and should be constrained from the phenomenological considerations. However, the significance of the value determined in \cite{Drummond:1979pp} comes from the fact that it can serve at least as a conservative lower limit for the value of $\tilde{q}$. Therefore, when discussing the potential observational implications we can take it as a reasonable assumption that the value of $\tilde{q}$ should be \textit{at least} of the order of magnitude predicted in \cite{Drummond:1979pp} or higher. In the following paragraph we will discuss the cosmological consequences of non-minimal coupling focusing on the case of this limit. Taking this conservative lower limit, $\tilde{q} = q_{3}/r_{0}^{2}$, where $r_{0}\sim2M$, to be of order 1, the mass $M$ has to be $\sim10^{13}$ kg, such that for most of the astrophysical sources of mass comparable to or higher than the solar mass, it is $\tilde{q} \ll 1$. Therefore, it follows that in such a case the observable effects will be negligible. 

However, an interesting candidate relevant even for this conservative limit is given by hypothetical astrophysical objects called Primordial Black Holes (PBHs) \cite{1967SvA....10..602Z, Carr:1974nx}, which have a mass range conjectured (depending on the model, i.e.~on the time when they formed after the Big Bang) to be between $10^{-8}$ kg and many solar masses. Another point to note is that Hawking's theory of black holes evaporation \cite{Hawking:1975vcx} predicts that any PBH of a mass less than $\sim10^{11}$ kg would have evaporated by now. Needles to say, it is of course very questionable to expect that this approximation -- which assumes a weak field and the one-loop approximation -- can be applicable for a PBH of mass $\sim10^{-8}$ kg ($\tilde{q}\sim10^{41}$). On the other hand, if we take the Drummond \& Hathrell \cite{Drummond:1979pp} theory to be applicable for $\tilde{q} \sim 1-10$ (i.e.~a mass of $\sim 10^{13}-10^{11}$ kg, since $\tilde{q} \propto 1/\sqrt{\rm mass}$), then it can be utilized for both kinds of PBHs -- the ones which are already evaporated and the ones which still exist. With regard to evaporated PBHs it is in principle quite likely that the local distribution of magnetic fields previously associated with them would still be present at their location. Thus, an astrophysical observation of a local suppression of the magnetic field can be further analyzed as a possible candidate for the detection of PBH. The investigation of PBHs in relation to non-minimal coupling can in general take two directions. In the case of a significant localized suppression of magnetic field being observed (while previously not being recognized as a PHB candidate), we can take non-minimal coupling considerations into account and claim this localized suppression to be a signature of a PBH. On the other hand, if some location is already claimed to be a PBH candidate by some other model or consideration, then -- if a significant suppression of magnetic field around this point is established -- it can serve as a way to establish a constraint on the value of $\tilde{q}$. For the PBHs which exist until the present day ($10^{11}$ kg $\lesssim$ mass $\lesssim 10^{13}$ kg) the same arguments as for the usual black holes in \cite{Pavlovic:2018idi,Pavlovic:2019rim} may be made. We note that the discussed question of PBHs is also interesting because it connects the question of non-minimal coupling with the question of the evolution of cosmological magnetic fields, which could also be connected with the question of the origin of the Universe \cite{Leite:2018bbo}. 

In any realistic modeling of an accretion disk around a black hole, the study of factors which affect the {\it isco} is important. The implications of the correct modeling of an accretion disk is also important for the analysis of the black hole images made by the Event Horizon Telescope \cite{EventHorizonTelescope:2019dse, EventHorizonTelescope:2019ths, EventHorizonTelescope:2019ggy} and for similar future observations. In the present work we saw that a non-minimal coupling affects {\it isco} not only quantitatively but also qualitatively. One of the means to distinguish between a black hole and hypothetical/exotic objects (like a naked singularity, a non-singular black hole, etc.) has been the investigation of the properties of {\it isco} in absence of and in presence of magnetic field (in the minimal coupling scenario). Including a non-minimal coupling considerations in these studies can answer the questions like -- can a black hole, when its gravity is non-minimally coupled with a magnetic field, have same/distinguishable observational signatures as those of exotic objects? For instance, the $\gamma$\,-\,metric -- one of the candidates for naked singularity -- shows two well separated regions of the {\it isco} for $\gamma \in (1/\sqrt{5},1/2)$ (in absence of \cite{Chowdhury:2011aa} and in presence of magnetic fields \cite{Benavides-Gallego:2018htf}). We can see from Fig.~\ref{iscoUni1left} that this particular feature is present for a black hole too when the non-minimal coupling is considered. So it would be interesting to further investigate how a black hole and exotic objects differ in their observational signature when non-minimal coupling of electromagnetic field with strong gravity is considered. Very often the estimation of spin of a black hole candidate has been derived from the theoretical relationship between spin and the {\it isco} radius \cite{Shafee:2005ef,Shafee:2007sa}. Though in the present work we have not considered a rotating black hole (i.e.~a Kerr black hole), from previous studies in minimal coupling we can expect that there too the properties of the {\it isco} would be different if non-minimal coupling is taken into consideration.

From observational astronomy it is well established that many astrophysical objects (active galactic nuclei, radio galaxies, both stellar mass and super-massive black holes, etc.) often have jets associated with them. Many proposed mechanisms for the formation of jets assumes the presence of a magnetic field in the system comprising a central object (often a neutron star or black hole), an accretion disk and a jet. The transfer of energy and material from accretion disk to jet is facilitated by magnetic fields in many models \cite{Blandford:1977ds}. So far only the minimal coupling of magnetic field with gravity have been considered in the models of jet formation. A study of non-minimal coupling of magnetic fields with gravity can possibly shed some new light on the nature of the phenomenon.

\begin{figure}[t]
\centering
\includegraphics[width=0.5\columnwidth]{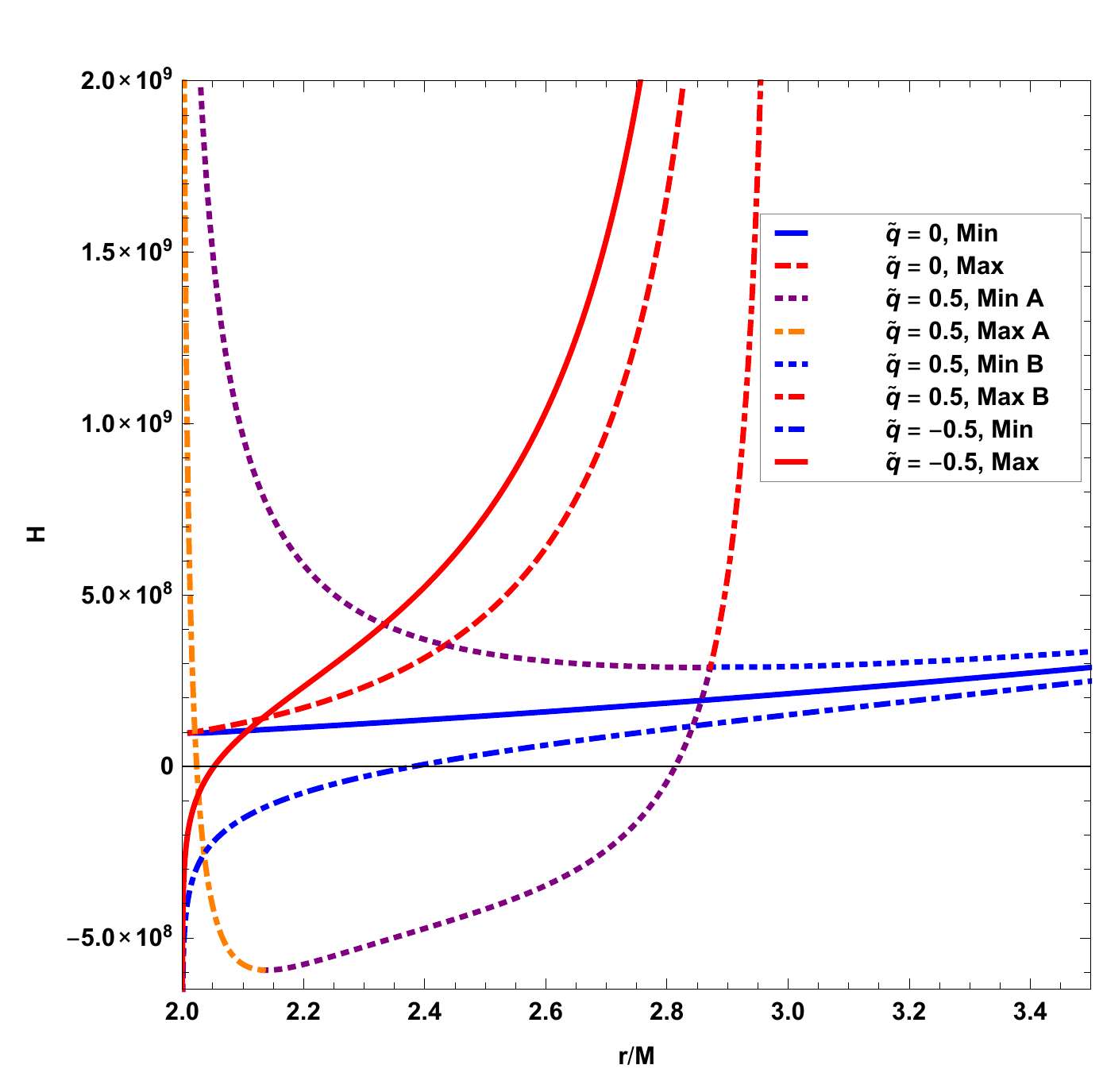}
\caption{The location of minima (denoted as ``Min") and maxima (denoted as ``Max") in the $r$-$H$ parameter space for different values of $\tilde{q}$. As discussed above, while for the case of minimal coupling and for $\tilde{q}<0$ only one minimum/maximum exists, for $\tilde{q}>0$ there are two minima/maxima. In this latter case ``Min A"/``Max A" denotes the minimum/maximum which is closer to the event horizon, while ``Min B"/``Max B" denotes the minimum/maximum farther away from the event horizon which converges with the one for $\tilde{q} \leq 0$ for large $r$ or $H$.}
\label{fig:minmaxplot}
\end{figure}

\begin{figure}[t]
\centering
\includegraphics[width=0.5\columnwidth]{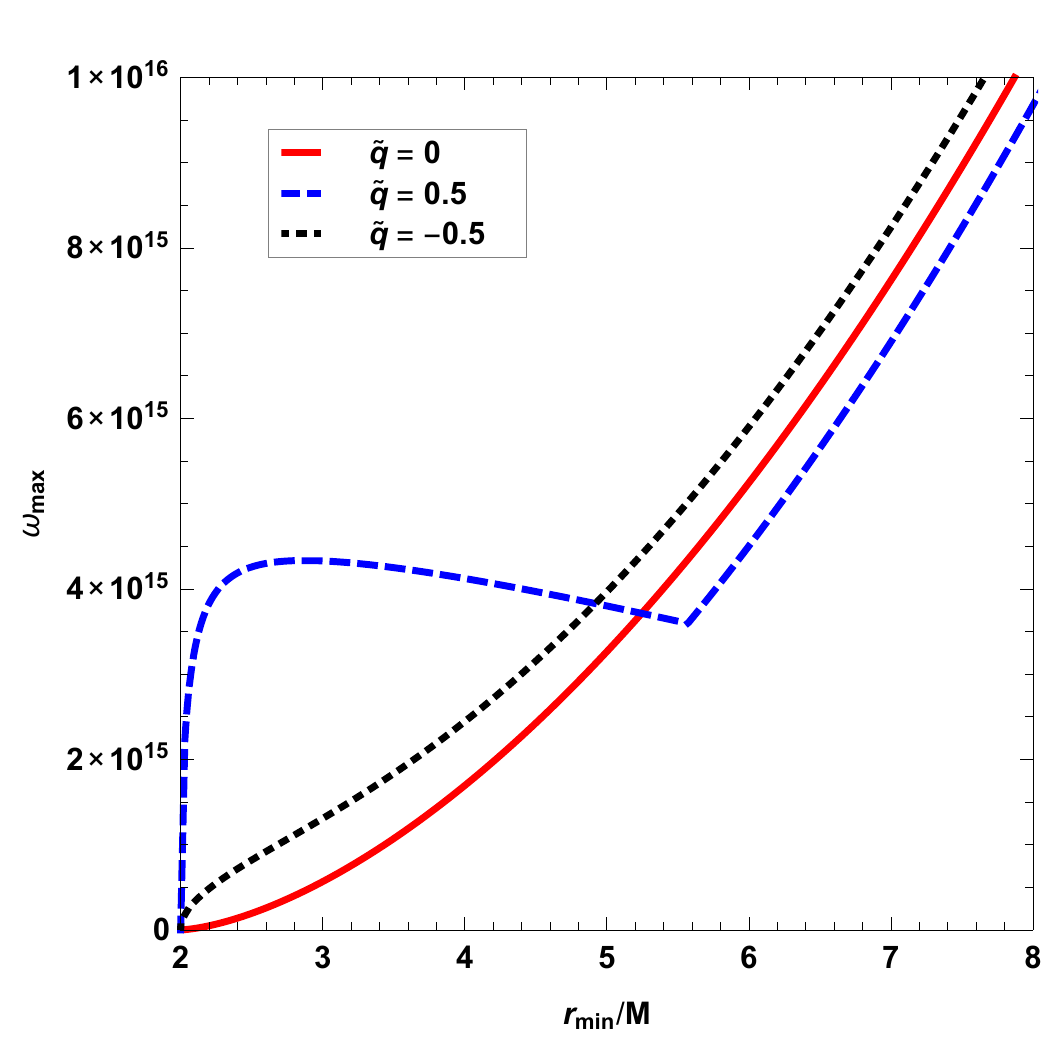}
\caption{Upper limit $\omega_{\rm max}$ for the frequency of the emitted synchrotron radiation by a proton orbiting a stellar black hole for different values of $\tilde{q}$ (according to Eq.~(\ref{eq:omegamax})) as a function of the radius $r_{\rm min}$ at which the minimum of the effective potential (and hence the stable circular orbit which the proton eventually reaches) is located. Note the deviation at small $r_{\rm min}$ for the case $\tilde{q} = 0.5$ which is due to the second minimum/maximum of the effective potential.}
\label{fig:omegamax}
\end{figure}

\subsection{Synchrotron Radiation} \label{sec:SynchRad}

Finally, as a concrete astrophysical application and in order to give a quantitative estimate how it may be changed by non-minimal coupling, we consider the possibility of synchrotron radiation from the vicinity of a black hole.
Doing so requires the trapping of high energy particles near the horizon \cite{Preti_2004}. In absence of a magnetic field the possibility of particle trappings within a radius of $6M$ in unbound and unstable orbits \cite{Davis:1972dm, Bardeen:1972fi, Misner:1972jf} has limitations. In the presence of a magnetic field, even in a minimal coupling scenario, highly energetic particles can be trapped in stable circular orbits within a radius of $6M$ and near the horizon. We saw here that non-minimal coupling further facilitates the trapping of high energy particles near the horizon, such that in any study of synchrotron radiation from the vicinity of a black hole the inclusion of non-minimal considerations can be useful.

In order to demonstrate a concrete possibility for such an investigation, we start with the concepts developed for synchrotron radiation from a weakly magnetized Schwarzschild black hole with minimal coupling in \cite{Shoom:2015uba} and extend them to the non-minimal case. In particular, we consider the case of a proton orbiting a stellar Schwarzschild black hole in a uniform magnetic field (cf.~Sec.~\ref{sec:UMF}) which translates to a value for $\beta$ of $\beta \simeq 4.718 \times 10^{7}$. 

In \cite{Shoom:2015uba}, for $\beta \gg 1$ and minimal coupling, the following relations for $H_{\rm max}(\beta,r)$ and $H_{\rm min}(\beta,r)$, the values for the location of the minimum and maximum of the effective potential in the parameter space, respectively, has been derived:
\begin{eqnarray}
H_{\rm max} &= \frac{\beta r_{\rm max}^{2}(r_{\rm max} - M)}{4 M^{2} (3 M - r_{\rm max})} + \mathcal{O}(\beta^{-1}) \,, \\
H_{\rm min} &= \frac{\beta r_{\rm min}^{2}}{4 M^{2}} + \frac{M}{2 \beta (r_{\rm min} - 2 M)} + \mathcal{O}(\beta^{-3})\,,
\end{eqnarray}
where $r_{\rm max}$ and $r_{\rm min}$ are the corresponding normalized radii. From this the respective energies $E_{\rm max}$ and $E_{\rm min}$ for a constant value of $H$ may be approximated by the expressions
\begin{eqnarray}
E_{\rm max}^{2} &\simeq \frac{\beta^{2} r_{\rm max} (r_{\rm max} - 2 M)}{(3 M - r_{\rm max})^{2}} \,, \\
E_{\rm min}^{2} &\simeq 1 - \frac{2 M}{r_{\rm min}} + \frac{M^{4}}{\beta^{2} r_{\rm min}^{3}(r_{\rm min} - 2 M)}\,.
\end{eqnarray}

Taking into account the form of the effective potential, this results in a helical trajectory around the circular orbit with radius $r_{\rm min}$. Without synchrotron radiation this helical movement would be stable, while due to the energy loss the particle ''spirals down'' to $r = r_{\rm min}$, such that in the end the trajectory is nearly circular (i.e.~it has descended to the minimum of the effective potential).

For this situation \cite{Shoom:2015uba} presents a well-motivated estimate for the upper limit $\omega_{\rm max}$ of the frequency of the synchrotron radiation, namely
\begin{equation} \label{eq:omegamax}
\omega_{\rm max} \simeq 0.87 \beta E_{\rm max} .
\end{equation}

When considering the non-minimal case, i.e.~$\tilde{q} \neq 0$, one first has to consider the structure of the effective potential, which can be done using Fig.~\ref{fig:minmaxplot}: For $\tilde{q} \leq 0$ we have a maximum and a minimum, however non-minimal coupling considerably shifts them in the $r$-$H$ parameter space and, furthermore, also regarding their actual value. This, in turn, results in a upwards shift of $\omega_{\rm max}$, as can be seen in Fig.~\ref{fig:omegamax}. This means that there is, in principle, a possibility to limit $\tilde{q}$ from below by measuring the synchrotron frequency emitted in the vicinity of a black hole.

A more interesting situation develops for $\tilde{q}>0$. As described above, in this parameter region, depending on the values of $r$ and $H$, up to two minima and maxima can exist. For this case we name the maximum closer to the event horizon, at $r_{\rm max}^{\rm A}$, ``maximum A" and the one farther from it, at $r_{\rm max}^{\rm B}$, ``maximum B", i.e.~$r_{\rm max}^{\rm A} < r_{\rm max}^{\rm B}$ (and analogous to that for the minima at $r_{\rm min}^{\rm A}$ and $r_{\rm min}^{\rm B}$). When calculating $\omega_{\rm max}$ in this case, one has to consider two different scenarios (as before, for a fixed $H$): $V_{\rm eff}^{2}(r_{\rm max}^{\rm A}) < V_{\rm eff}^{2}(r_{\rm max}^{\rm B})$ and $V_{\rm eff}^{2}(r_{\rm max}^{\rm A}) \geq V_{\rm eff}^{2}(r_{\rm max}^{\rm B})$. In the former scenario, in order to have a conservative limit, a particle which eventually ends up in the minimum A can initially at most have an energy corresponding to the maximum A, while a particle arriving at the minimum B can start with an energy corresponding to the maximum B. On the other hand, in the second scenario, particles finishing at the trajectories corresponding to either of the two minima may start at the higher energy, i.e.~at the one corresponding to maximum A. This, together with the particular values of the effective potential at the maxima, results in the rather different behavior of $\omega_{\rm max}$ with $r_{\rm min}$ for $\tilde{q}=0.5$ in Fig.~\ref{fig:omegamax}, as for very small values of $r_{\rm min}$ close to the event horizon the maximum frequency is dramatically higher than for the case of $\tilde{q} \leq 0$, which might result in either a clear observational feature or, in case it is not observed, a rather strict upper limit on $\tilde{q}$.

\section{Concluding Remarks} \label{sec:Conclusions}

New astrophysical observations, such as the results obtained by the Event Horizon Telescope and the study of super-massive black holes, make it possible to investigate the structure of magnetic fields near the event horizon more precisely \cite{EventHorizonTelescope:2021srq, Ricarte:2021frd, Eatough:2013nva, Piotrovich:2020ooz}. This topic is also of fundamental theoretical interest since black hole horizons are associated with strongest gravitational fields currently accessible to observations. It is precisely in this strong gravity setting that one could expect to find some signatures of new effects going beyond the standard description of relationship between gravity and electromagnetism. As we have discussed, one of such effects -- that can be motivated from the fundamental field-theory considerations -- is the non-minimal coupling of gravity and electromagnetism. For instance, this type of non-minimal coupling naturally comes as a consequence of the effect of QED vacuum polarization on curved spacetime. We note that in the setting of Schwarzschild spacetime near the horizon one could also expect other effects beyond standard electrodynamics to manifest, such as some non-linear effects \cite{Bokulic:2019kcc}. With an aim to study this effect in a setting which is relevant for the astrophysical application, we have considered the case of non-minimal coupling between magnetic fields and gravity on Schwarzschild spacetime in the weak magnetic field limit. We have thus followed, improved and considered some new applications of the ideas recently proposed in \cite{Pavlovic:2018idi,Pavlovic:2019rim}, with the primary motivation to elaborate the physical consequences of the non-minimal coupling between gravity and electromagnetism and to help bringing them closer to the possibility of observational verification.

In the following we are going to summarize our findings and the conclusions which may be drawn from them.

\begin{itemize}

\item We have first reviewed the already known solutions for the case of minimally coupled magnetic fields on Schwarzschild spacetime and then considered the non-minimal coupling effects for both the asymptotically dipole and uniform magnetic fields in detail, while discussing the changes induced by the non-minimal coupling with respect to the minimal coupling case. We have also provided a detailed discussion of the changes that come as a result of non-minimal coupling in both of the mentioned magnetic field configurations. In all of the discussed settings we have found that even the modest values of the coupling parameter $\tilde{q}$ -- which do not seem to appear to be in contradiction with other observations -- can cause a significant change in the magnetic fields near the event horizon. These changes can have the character of an amplification or a suppression, depending on the sign of the non-minimal coupling parameter.

\item Apart from this quantitative change (which appears for both the asymptotically dipole and the asymptotically uniform case), in the case of an asymptotically uniform field a further qualitative change also appears: the direction of magnetic fields changes near the horizon -- for the positive value of $\tilde{q}$ the direction alters near the equatorial plane, while for the positive values the direction alters near the two poles.

\item We have also studied in detail how non-minimal coupling modifies the features of the effective potential in the equatorial plane. The non-minimal coupling can increase/decrease the effective potential, change the position of the potential well and raise or lower the level of the maxima and minima. Due to these reasons, under proper conditions non-minimal coupling can significantly change the spectrum of the scattered particles, affect the innermost stable circular orbit (\textit{isco}), influence the energy spectrum of trapped particles and change the center of mass energy during the acceleration of charged particles. All these processes were discussed in detail qualitatively and quantitatively. The most promising effect for manifesting the difference between minimal and non-minimal coupling seems to be the study of the spectrum of scattered particles for the case of a dipole field configuration.

\item We have also discussed various physical scenarios of observational interest in which the non-minimal coupling effects could play a significant role. Such scenarios and settings include primordial magnetic fields and primordial black holes, studies of the consequences for the \textit{isco} (which, for instance, are of interest for discrimination between black holes and potential exotic objects with similar properties like non-singular black holes, naked singularities, etc.), potential effects of non-minimal coupling in jet formation and synchrotron radiation.

\item For the latter we analyzed the potential change of the observational signatures due to non-minimal coupling in more detail, such that in the future it might be used to set limits on the value of $\tilde{q}$. In particular, for non-minimal coupling the frequency of the radiation emitted at an intermediate distance from the black hole might be moderately shifted up or down for negative and positive values of $\tilde{q}$, respectively, compared to the minimal case. On the other hand, close to the Schwarzschild radius a positive $\tilde{q}$ might result in a dramatic increase of the (maximal) synchrotron frequency, hence potentially giving the most significant observational signature.

\end{itemize}

We believe that future theoretical research done in these topics -- combined with further improvements in experimental techniques for the study of magnetic fields near the black hole horizon -- can lead to the potential observation of non-minimal coupling, or at least strongly constraining the value of the coupling parameter. Such progress could be a significant step forward in our understanding of strong gravitational fields and quantum effects on them, as well as the connection between electromagnetism and gravity. In fact, if observed, non-minimal coupling between electromagnetism and gravity would be the first detection of a curved spacetime quantum process.

\section*{Acknowledgments}

The work of AS is supported by the Russian Science Foundation under grant no.~22-11-00063.

\section*{Data Availability Statement}

The data generated and/or analyzed during the current study are not publicly available for legal/ethical reasons but are available from the corresponding author on reasonable request. The data that support the findings of this study are available upon reasonable request from the authors.

\section*{References}

\providecommand{\newblock}{}

\end{document}